\documentclass[structabstract]{aa}  
\usepackage{graphicx}
\usepackage{txfonts}
\usepackage{upgreek}
\usepackage{tipa}
\usepackage{verbatim}
\usepackage{natbib}
\usepackage{color}

\newcommand{\hii}{H{\small{II}}}

\newcommand{\lsol}{L$_\odot$\,}
\newcommand{\msol}{M$_\odot$\,}
\newcommand{\point}{\cdot}
\newcommand{\kms}{km$\cdot${s$^{-1}$}}
\newcommand{\kkms}{K$\cdot$km$\cdot${s$^{-1}$}}

\newcommand{\amin}{$^{\prime}$}                   %arcus and coordinates
\newcommand{\asec}{$^{\prime \prime}$}
\newcommand{\adeg}{$^{\circ}$}

\def\kms{km\thinspace s$^{-1}$}

\begin{document}
\newcommand{\cf}{\textit{see}}
   \title{S-bearing molecules in Massive Dense Cores}

   \subtitle{}

   \author{F. Herpin\inst{1,2} \and M. Marseille\inst{3} \and V. Wakelam\inst{1,2} \and S. Bontemps\inst{1,2} \and D.C. Lis\inst{4}}

   \institute{Universit\'e de Bordeaux, Laboratoire d'Astrophysique de Bordeaux, F-33000 Bordeaux, France
   \and
    CNRS/INSU, UMR 5804, BP 89, 33271 Floirac cedex, France.\\
              \email{herpin@obs.u-bordeaux1.fr}
   \and
   SRON Netherlands Institute for Space Research, Landleven 12, 9747AD Groningen, The Netherlands. \\           
   \and 
   California Institute of Technology, Downs Laboratory of Physics 320-47, Pasadena, CA 91125.
   }

    \offprints{F. Herpin}
  \date{Received October 30th, 2008; accepted May 28th, 2009}

% \abstract{}{}{}{}{} 
% 5 {} token are mandatory
 
  \abstract
{High-mass stars, though few
in number, play a major role in the interstellar energy budget and the
shaping of the Galactic environment. However, despite this importance, the formation of high-mass stars is not well
understood, due to their large distances, short time scales, and heavy extinction.
}
% Aims
{Chemical composition of the massive cores forming high-mass stars can put some constrains on the time scale of the massive star formation: sulphur chemistry is of specific interest due to its rapid evolution in warm gas and because the abundance of sulphur bearing species increases significantly with the temperature. }
% Methods
{Two mid-infrared quiet and two brighter massive cores are observed in various transitions (E$_\mathrm{up}$ up to 289~K) of CS, OCS, H$_2$S, SO, SO$_2$ and of their $^{34}$S isotopologues at mm wavelengths with the IRAM 30m and CSO telescopes. 1D modeling of the dust continuum is used to derive the density and temperature laws, which are then applied in the RATRAN code to model the observed line emission, and to derive the relative abundances of the molecules.
}
% Results heading (mandatory)
{All lines, except the highest energy SO$_2$ transition, are detected. Infall (up to 2.9~\kms) may be detected towards the core W43MM1. The inferred mass rate is 5.8-9.4~10$^{-2}$~M$_{\odot}/$yr. We propose an evolutionary sequence of our sources (W43MM1$\rightarrow$IRAS18264$-$1152$\rightarrow$IRAS05358$+$3543$\rightarrow$IRAS18162$-$2048), based on the SED analysis. The analysis of the variations in abundance ratios from source to source reveals that the SO and SO$_2$ relative abundances increase with time, while CS and OCS decrease.
}
{Molecular ratios, such as [OCS/H$_2$S], [CS/H$_2$S],  [SO/OCS], [SO$_2$/OCS],  [CS/SO] and [SO$_2$/SO] may be good indicators of evolution depending on layers probed by the observed molecular transitions. Observations of molecular emission from warmer layers, hence involving higher upper energy levels are mandatory to include. 
}


   \maketitle
%
%________________________________________________________________

\section{Introduction}
	
%on pense que l'Žvolution des massives dense cores est la meme chose qu'Žtudier l'Žvolution d'un proto cluster.
%dŽmontrer que regarder ˆ cet Žchelle est dŽterminante (accrŽtion compŽtitive, coalescence...).
%Leurini, la masse est concentrŽe sur une sous-source qui domine.}

OB stars are the main contributors to the evolution and energy budget
of galaxies. Their formation, however, is not understood yet and the "classical" scheme for low-mass star formation \citep[see][and references therein]{andre2000} cannot be applied as such to OB stars. Indeed, young OB stars and protostars strongly interact with the surrounding massive clouds and cores leading to a complex and still not clearly defined sequence of objects from pre-stellar cores which are often believed to be hosted in the so-called IR dark clouds (IRDCs), to High-Mass Protostellar Objects (HMPOs), to hot cores and UItra Compact HII regions \citep[e.g.][]{beuther2007b,menten2005}.

 This tentative evolutionary scheme is however mostly based on observations of distant objects, for which, the physical scales are often at the order of 0.5 pc. In contrast, the well established protostellar evolution from Class 0 to Class III Young Stellar Objects (YSOs) for low-mass stars revealed that the envelopes of individual protostars have typical sizes ranging from 0.02 to 0.04 pc in clustered star-forming regions \citep[see][and references therein]{motte2001}. HMPOs are therefore more protoclusters than individual protostars. 
Recent interferometric observations have confirmed that they are fragmented and that they actually form clusters of stars
 \citep[e.g.][]{beuther2004,shepherd2003}. More recently,
 \citet{motte2007} have obtained a complete view of high-mass
 protostellar phases at the scale of a giant molecular cloud which is
 not too distant (Cygnus X at 1.7 kpc). This complex is rich enough to
 provide a first reliable statistics (40 massive dense cores) at a
 physical scale of $\sim$ 0.1 pc. From this unbiased survey of the
 earliest phases of the high-mass star formation, it was deduced that
 massive pre-stellar dense cores are extremely rare, and that of the
 order of 50 \% of the massive dense cores are already forming
 high-mass stars while being still cold and not bright in the IR
 (IR-quiet dense cores). The remaining 50 \% of dense massive cores  can be referred as IR-bright dense cores. Referring to the typical sizes discussed in \citet{williams2000}, these 0.1 pc size objects are dense cores, in contrast to clumps which are more 1 pc size. On the other hand, these 0.1 pc dense cores are not necessary precursors of single stars and may form a group of stars, and would therefore be considered as clumps in the primary definition of   \citet{williams2000}. 
We will here follow the scheme discussed in \citet{motte2008} in which the study of massive dense cores at physical scales of $\sim$ 0.1 pc or less is required to really address the question of evolutionary stages  for high-mass star and cluster formation. The interest of studying the evolution of massive dense cores is also supported by the most recent theoretical scenarios, which predict either a monolithic collapse of turbulent gas at the scale of massive dense cores \citep[][]{mckee2002,  krumholz2008}, or a competitive accretion inside the gravitational potential of a cluster-forming massive dense core \citep{bonnell2006}. Finally, it is worth noting that some direct observational evidence for the so-called global collapse has usually been obtained for massive dense cores at the typical $\sim$ 0.1 pc scale \citep[e.g.][]{motte2005}.
 
 %New massive star forming regions have been discovered \citep[\textit{e.g.} in Cygnus X or in W43;][]{motte2007, motte2003} and details %on the structure of massive protostellar envelopes are emerging. Indeed, r
 The chemical composition of the
protostellar envelopes can be used to constrain the ages of low-mass protostars
\citep[][]{millar1997,hatchell1998,doty2002,wakelam2004a}. The sulphur
chemistry is of specific interest because of its rapid evolution in
warm gas and since the abundances of sulphur bearing species increase
significantly with temperature, both by ice evaporation and by
shock interaction. Sulphur could thus act as a clock on time scales relevant to the embedded phase of star formation \citep[][]{charnley1997,wakelam2004a,wakelam2004b}. 
%In fact, the molecular abundance sensitivity with evolutionary state and physical parameters in low-mass stars was poorly understood before  %\citet{wakelam2004a} and  \citet{wakelam2004b}. 
A good sketch is now established for these objects allowing to date them within a ``class'' of objects. Here we propose some first constraints on the time of evolution of the targeted massive cores; detailed chemical modelling will be presented in a forthcoming paper (Wakelam et al., in preparation). 
 
For high-mass protostars, the first studies using sulphur bearing molecules as evolutionary tracers are reported in \citet{hatchell1998} and \citet{vandertak2003}. We present here a complementary
  study of sources, which are weaker at mid-infrared wavelengths. Also in contrast to \citet{vandertak2003} who used chemical models without the inclusion of atomic oxygen  \citep[e.g.,][]{charnley1997}, we prefer to use the recent results of \citet{lis2001} and \citet {vastel2002}. Furthermore, \citet{wakelam2004b} underlined that the sulphur may be mainly in the atomic form in hot cores.  

 To analyze these observations, we will apply a method similar to
 \citet{wakelam2004a} who measured the age of the hot corinos of
 IRAS16293$-$2422 with reasonable success. Our observations of low-
 and high-energy transitions for each selected molecule enable us to
 probe the inner ($>$ 100~K) and outer ($<$ 100~K) regions of the
 objects studied. Using transitions at several energies, we first
 constrain, with a radiative transfer model
 \citep[RATRAN,][]{hogerheijde2000}, the sulphur bearing abundance
 profiles within the envelopes (this paper). In a forthcoming paper,
 the derived abundance profiles will be compared to a grid of chemical models (NAHOON) adapted to the physical structure of each source in order to constrain the age of each source. To cover the complete range of energy of the massive envelopes  \citep[from 20~K to more than 100~K,][]{doty2002}, transitions with upper energies between 10 and 110~cm$^{-1}$ are required. Indeed, even if observing emission lines covering a large energy range is mandatory to derive the excitation conditions of the molecules,  the only way to probe the inner warmer regions is to observe high energy transitions which levels are not populated in the cold outer gas. Actually, probing the inner region of massive dense cores is the only way to chemically date their evolution from the molecular cloud.

\section{Description of the Source Sample}
\label{source}

 \begin{table}[t]
\begin{minipage}[t]{\columnwidth}
%\begin{flushleft}
 \caption{\label{sources} Source Sample. The $F_{21}$ value is the MSX value for the 21 $\mu$m flux, corrected for the distance of each source ($F_{21}\times(d/1.7)^2$). }
\label{table1}      % is used to refer this table in the text
\resizebox{\hsize}{!}{   
\begin{tabular}{l lll ccc}
  \hline
\noalign{\smallskip}
            &   \multicolumn{2}{c}{Source Coordinates\footnote{Coordinates correspond to the peak of the mm continuum emission from \citet{beuther2002c} and \citet{motte2003},  respectively for IRAS05358+3543, IRAS18264$-$1152 and W43MM1. For IRAS18162-2048, coordinates correspond to the CO emission peak from \citet{benedettini2004}.}}    &   
& \multicolumn{2}{c}{Properties\footnote{Luminosity, distance and velocities are from \citet{sridharan2002} (IRAS05358$+$3543 and IRAS18264$-$1152),  \citet{benedettini2004} (IRAS18162$-$2048) and \citet{motte2003} (W43MM1).}} & $F_{21}$ \\
Source     & RA  (J2000.0) & Dec  & $V_{\rm LSR}$  & $L_{\rm bol}$ & $d$  \\
\noalign{\smallskip}
&   ( h m s )        &( \adeg\ \amin\ \asec\ )& (km s$^{-1}$) &($L_\odot$) 
& (kpc) & (Jy) \\
\noalign{\smallskip}
\noalign{\smallskip}
\hline
\noalign{\smallskip}
\noalign{\smallskip}
\noalign{\smallskip}
\noalign{\smallskip}
IRAS05358+3543      & 05 39 13.1    &   +35 45 50   & $-$17.6   & $6.3\,\times\,10^3$     & 1.8  &  18.9 \\
IRAS18162$-$2048    & 18 19 12.1    & $-$20 47 31   & +12.1     & $2.0\,\times\,10^4$     & 1.9   &  141.4 \\
IRAS18264$-$1152    & 18 29 14.4    & $-$11 50 23   & +43.6     & $1.\,\times\,10^4$     & 3.5 &  4.9 \\
W43MM1             & 18 47 47.0    & $-$01 54 28   & +98.8     & $2.3\,\times\,10^4$     & 5.5   & 4.8 \\
\noalign{\smallskip}
\hline
\end{tabular}
  }
%\end{flushleft}  
\end{minipage}
\end{table}

    \begin{figure}
   \centering
   \includegraphics[width=\columnwidth]{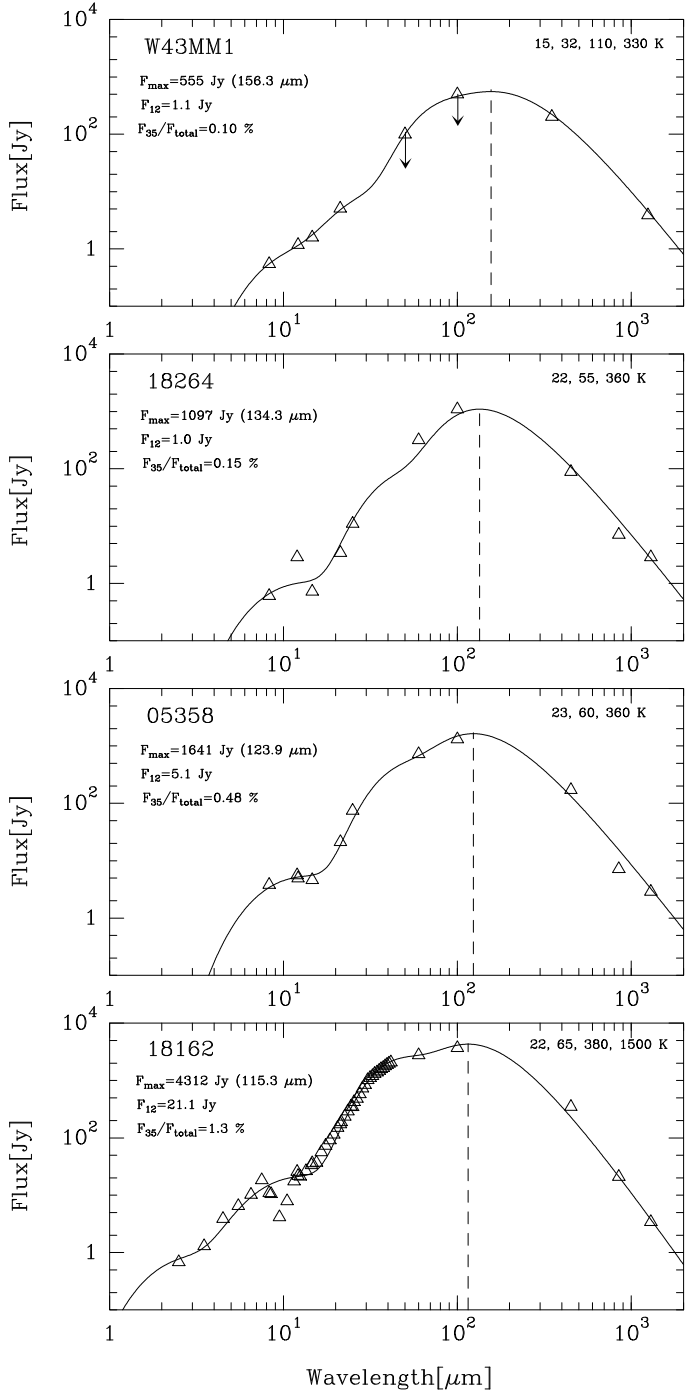}
   \caption{Spectral energy distributions obtained from 1D model overlaid on
      observed fluxes of each source. Peak fluxes in the millimeter range have been adjusted to fit radial extension of the model. In each caption are given the temperatures of the black body components (top right), the maximum flux $F_\mathrm{max}$  and the corresponding wavelength of the SED, the flux $F_{12}$ at 12$\mu$m, and the contribution of the hot part ($F_{35}$, integrated flux for $\lambda <$35 $\mu$m) to the total integrated flux $F_\mathrm{total}$ (top left).}
              \label{Figsed}%
    \end{figure}

The selected sources are two IR-quiet dense cores \citep[MSX Flux$_{21\mu m}<10$ Jy at 1.7 kpc, following the definition of][]{motte2007}  and two slightly brighter ones (IRAS05358+3543 and IRAS18162$-$2048), with bolometric luminosities 0.6--2.3~$10^4$~L$_{\odot}$ at distances 1.8--5.5~kpc (see Table \ref{sources}) and sizes of $\sim$0.11--0.13~pc (as seen in the beam of our observations, see Sect.~\ref{seddet}). These massive dense cores are expected to be fragmented at small scales and are meant to represent the earliest phases of the formation of clusters. The expected multiplicity can however be neglected to first order, to follow the average chemical evolution of their surrounding dense cores. In fact, a small fraction of the total mass of the cores is in the individual fragments, and usually the large scale average density profile of the cores applies well down to the smallest observed structures. This suggests that the cores could be well represented to first order as spherically symmetric structures from a few 0.1 pc down to $\sim 0.01$ pc. IRAS05358+3543 has for instance been studied with millimeter interferometers  \citep[see][]{leurini2007}. It appears that the $\sim 0.3$ pc size core of 613$\,$M$_\odot$ splits into two main objects (mm1a and mm1b) of 1 and 0.6$\,$M$_\odot$ inside regions of 0.006 pc diameter, separated only by $\sim 0.008$ pc. A mass of 1.6$\,$M$_\odot$ in a region $\sim 0.01$ pc comprising mm1a and mm1b, inside a 0.3 pc size core of 613$\,$M$_\odot$ corresponds to an average radial profile of the density of $r^{-1.25}$ which is actually close to 
the $r^{-1.4}$ profile derived from large scale by
\citet{beuther2002a} and which we use to describe the global 1D distribution of matter in the core. Two other millimeter condensations mm2 and mm3 are also detected with the interferometer. They could 
eventually confuse the results for molecular tracers which can trace such high density regions. 

A large number of continuum and molecular line (CO, SiO, CH$_3$OH,
H$_2$CO etc) data towards these sources have been published
\citep[\textit{e.g.}][]{benedettini2004}. The physical conditions have
thus already been determined in some of the envelopes
\citep[\textit{e.g.}][]{beuther2002a} and we use the published data
together with our observations. We limit our study to massive
  dense cores to simplify this first step of the analysis for two
  reasons. First, different physical conditions have been observed in
these objects, but they still have the same structure in the sense
that there is no strong UV field or ultra-compact \hii\ regions, which
will change drastically the chemistry with time. The second reason is that these cores could be for massive dense cores the equivalent of the Class 0 YSOs in the general scenario of gas and dust evolution, for which sulphur chemistry \citep[][]{wakelam2004a,wakelam2004b} has already been studied. Finally, the
sources studied here are part of the sample that will be studied within the key program {\em WISH}\footnote{http://www.strw.leidenuniv.nl/WISH/, PI: E. van Dishoeck}  with the HIFI instrument aboard the ESA Herschel Space Observatory.  

Both IRAS05358+3543 and IRAS18264$-$1152 are part of the
\citet{sridharan2002} sample. IRAS05358+3543  is a relatively low
luminosity ($L \sim6.3~10^3$~\lsol) and nearby (1.8~kpc) massive dense
core, composed of three main sources (mm1, mm2 and mm3; total mass
around 613~\msol), all within 4\arcsec $\times$9\arcsec\
\citep[][]{leurini2007}, hence within the telescope beam. The source
exhibits a hypercompact \hii\ region which may evolve into an UC-\hii\
region. \citet{beuther2007} show that the source actually splits into
four individual continuum components, two of which are part of a
proto-binary system, mm1 (dynamical age $\sim3.6~10^4$~yr). According
to \citet{longmore2006}, the main mid-IR source coincides with mm1a of
\citet{beuther2007}, but the second mid-IR source does not
  coincide with the mm source mm1b, hence suggesting even more sub-structure. The source
exhibits a class II methanol maser indicating that it is a very young
object. The source mm3 is the coldest and youngest
(T$_\mathrm{dust}\leq 20$~K). Several molecular outflows are observed
\citep[two from mm1, outflow rate of $6 \times 10^4$~\msol;][]{beuther2002c}. Strong line emission asymmetries in HCO$^+$ have been observed by \citet{fuller2005}.

IRAS18264$-$1152 is a massive dense core at a distance of 3.5~kpc. Its luminosity is estimated to be up to $10^4$~\lsol. A massive molecular outflow is centered on the mm dust peak \citep[][]{beuther2002c}. 
Continuum observations with Plateau de Bure \citep[][]{qiu2007} reveal that the source is double; the mass of the dust-gas core is estimated to be 570~\msol, while the outflow mass is 20.2~\msol (dynamical age of 0.5~$10^4$~yr). 

 \citet{benedettini2004} performed multi-line observations of IRAS18162$-$2048, a luminous infrared source  \citep[L$=2$~$10^4$~\lsol;][]{yamashita1989} at a kinematic distance of 1.9~kpc  \citep{kurtz1994}. This object is associated with the IRAS source and is actually an IR-bright dense core. A very powerful outflow ($M\sim$570~\msol), one of the most massive known, was observed (dynamical age $\sim$10$^6$~yr). The source has been resolved into a cluster of several objects  \citep[][]{stecklum1997}.

W43 is a massive star-forming region, located at 5.5~kpc, which harbors a giant \hii\ region and was studied in detail by \citet{motte2003}. W43MM1 is the most massive dense core of the region and does not coincide with any known IR source. The dust temperature is estimated to be 19~K and its luminosity  $2.3$~$10^4$~\lsol. W43MM1 harbors a methanol and a water maser, but does not show any infrared or centimeter emission. An infall has been detected by Motte et al. (private communication).

Continuum emission from our sources is quite well determined with well sampled observations from IRAS, MSX, SCUBA/JCMT or SHARC/CSO \citep[][]{hunter2000,jenness1995,minier2005}, SMA \citep[][]{beuther2007,su2004} and MAMBO/IRAM30m \citep[][]{beuther2002a}. The SED for IRAS18162$-$2048 is particularly well constrained, thanks to a complete ISO-SWS observation. Fewer continuum observations are available for the source W43MM1.

From the fitted SEDs (see Sect. \ref{seddet}), shown in Fig.~\ref{Figsed}, we make a rough evolutionary classification of our 4 objects, using the following parameters:
\begin{itemize}
  \item wavelength and flux of the maximum continuum emission,
  \item flux at 12 $\mu$m,
  \item contribution of the hot part ($\lambda <$35 $\mu$m) to the total integrated flux,
  \item temperature of the black-body components.
\end{itemize}
We assume that the less evolved the source is, the coldest it is, hence the SED peaks at longer wavelength with weaker flux. As the massive core evolves, it becomes warmer hence heating the dust wherein it is embedded: as a consequence the contribution of the flux at shorter wavelength ($F_{35}$, the integrated flux for $\lambda <$35 $\mu$m) increases.

Applying these criteria, W43MM1 appears to be the youngest object. A sequence of evolution W43MM1~$\rightarrow$ IRAS18264$-$1152~$\rightarrow$ IRAS05358+3543~$\rightarrow$ IRAS18162$-$2048 is proposed and adopted for the following discussion. This evolutionary sequence is also suggested by the dynamical ages derived by the authors cited above.

\section{Observations}

\begin{table*}[t!]
\begin{minipage}[t]{500pt}
\caption{\label{obs} List of observational parameters for the IRAM 30m
  and CSO telescopes (observed species, energy level transitions, upper energy, line emission frequencies, half power beam width, instrument, main beam efficiency $\eta_\mathrm{mb}$, receiver name, velocity resolution $\delta \varv$ and system temperature $T_\mathrm{sys}$.}           % title of Table
\label{table2}      % is used to refer this table in the text
\renewcommand{\footnoterule}{}
\begin{center}
\begin{tabular}{lccccccccc}
\hline
\hline
Species 	& Transition	& E$_{\rm up}$ &Frequency	& HPBW	& Instrument$^a$	& $\eta_{mb}$ & Receiver	& $\delta \varv$     & $T_{sys}$\\ 
		&		 	&  [cm$^{-1}$] &[GHz] 		& [\arcsec]		&         &      &   & [m$\point$s$^{-1}$]  &        [K]             \\
\hline
CS				& $3-2$ & 10 & 146.969 & 17 & 30m & 0.69 & D150 &  160  & 315-350  \\
				& $5-4$ & 24 & 244.936 & 10 & 30m & 0.48 & A230 &  96  & 530-860 \\
				& $7-6$ & 46 & 342.883 &  24 & CSO & 0.75 & 345 & 43 &  1210-1340 \\
\hline
C$^{34}$S			& $3-2$ & 10  & 144.617 & 17 & 30m & 0.69 & D150 &  162  & 270-390 \\
				& $7-6$ & 46 & 337.396  & 25 & CSO & 0.75 & 345 & 43 & 1260-1760  \\
\hline
SO 				& $3_4-2_3$ & 11 & 138.178 & 18 & 30m & 0.70 & D150 & 170  & 300-330  \\
				& $5_6-4_5$ & 24 & 219.949 & 11 & 30m & 0.55 & B230 & 110 & 590-930   \\
 				& $6_5-5_4$ & 35 & 251.825 & 10 & 30m & 0.48  & D270 & 190 & 750-830 \\
 				& $8_8-7_7$ & 61 & 344.308 & 24 & CSO & 0.75 & 345 & 43 & 1020-1050 \\
\hline
$^{34}$SO 		& $3_4-2_3$ & 11 & 135.775  & 18 & 30m & 0.70 & D150 & 170 & 290-350 \\
 		& $8_8-7_7$ & 61 & 337.582  & 25 & CSO & 0.75 & 345 & 43 & 1020-1060   \\ 
\hline
SO$_2$ 			& $5_{1,5}-4_{0,4}$ & 11 &135.696 & 18 & 30m & 0.70 & D150 &170  & 270-400   \\
 			& $10_{0,10}-9_{1,9}$ & 34 & 160.827 & 15 & 30m & 0.66 & C150 & 150 & 350-520  \\
 			& $11_{1,11}-10_{0,10}$ & 42 &221.965 & 11 & 30m & 0.55 & B230 & 100 & 530-550 \\
 			& $14_{0,14}-13_{1,13}$ & 65 & 244.254 & 10 & 30m & 0.48 & A230 & 190 & 830-1430   \\
 			& $18_{0,18}-17_{1,17}$ & 105 & 321.330 & 26  & CSO & 0.75 & 345 & 45 & 770-820   \\ 
			& $28_{4,24}-28_{3,25}$ & 289 & 267.719 & 9  & 30m & 0.46 & D270 & 170 & 470-510 \\  \hline
$^{34}$SO$_2$ 	& $5_{1,5}-4_{0,4}$ & 11 & 133.471 & 18 & 30m & 0.70 & D150 & 170 & 200-220 \\
\hline
OCS 			& $8-7$ &  14 & 97.301 & 25 & 30m & 0.77 & A100 & 240 & 140-160   \\
 			& $13-12$ & 37 & 158.107 & 15 & 30m & 0.66 & C150 & 150 & 340-360   \\
 			& $19-18$ & 77 & 231.060 & 11 & 30m & 0.52 & A230 & 100 & 450-580  \\
\hline
OC$^{34}$S 		& $8-7$ & 14 & 94.922 & 25 & 30m & 0.77 & B100 & 240 & 140-160 \\
\hline
H$_2$S 			& $1_{1,0}-1_{0,1}$ & 19 & 168.762 & 14 & 30m & 0.70 & D150 & 140 & 610-800  \\
 			& $2_{2,0}-2_{1,1}$ & 58 & 216.710 & 11 & 30m & 0.57 & B230 & 110 & 410-600  \\
\hline
H$_2$$^{34}$S 	& $1_{1,0}-1_{0,1}$ & 19 & 167.910 & 14 & 30m & 0.70 & D150 & 140 & 630-670 \\
\hline
& & & & &  \\
\end{tabular}\\
\end{center}
$^a$ Conversional factor is $S/T_{mb}=4.95$~Jy/K for IRAM 30m telescope and $S/T_{mb}=74$~Jy/K for CSO telescope.
\end{minipage}
\end{table*}
The observations presented here were performed during two sessions.
The first one in September 2005, using the IRAM-30m
antenna\footnote{http://iram.fr/IRAMES/index.htm}. The second one in
May 2006, using the Caltech Submillimeter
Observatory\footnote{http://www.cso.caltech.edu. The
  Caltech Submillimeter Observatory is supported by the National
  Science Foundation under award AST-0540882.}, on the summit of Mauna Kea (Hawaii). 

With the 30m telescope, all sources were observed in the rotational transitions lines of CS\,$J$=5$-$4, SO\,$J$=$3_4-2_3$, $J$=$5_6-4_5$, $J$=$6_5-5_4$, SO$_2$\,$J$=$5_{1,5}-4_{0,4}$, $J$=$10_{0,10}-9_{1,9}$, $J$=$11_{1,11}-10_{0,10}$, $J$=$14_{0,14}-13_{1,13}$, OCS\,$J$=$8-7$, $J$=$13-12$, $J$=$19-18$, H$_2$S\,$J$=$1_{1,0}-1_{0,1}$, $J$=$2_{2,0}-2_{1,1}$, and in the isotopic lines of $^{34}$SO\,$J$=$3_4-2_3$, SO$_2$\,$J$=$5_{1,5}-4_{0,4}$, OC$^{34}$S\,$J$=$8-7$, H$_2^{34}$S\,$J$=$1_{1,0}-1_{0,1}$ (see Table~\ref{obs}). Only IRAS05358$+$3543, IRAS18264$-$1152 and W43MM1 were observed in the CS and C$^{34}$S\,$J$=$3-2$ lines.  Observations were made using multiple receivers (either simultaneously A100, B100, A230, B230 or C150,  D150, C270, D270) coupled to the high-resolution VESPA backend (resolution of 80~kHz) in frequency switching model  (frequency throw of 7.9 MHz). The sky conditions were reasonable ($\tau_{0}^\mathrm{atm} \sim$0.1-0.5 at 225 GHz, see Table~\ref{table2} for the T$_\mathrm{sys}$). The telescope pointing (better than 2") and focus have been set on suitable planets (\textit{i.e.} Jupiter) or standard calibration sources.

Complementary observations have been performed with the CSO telescope for lines with frequencies higher than 300~GHz (in order to better cover the energy ladder). As IRAS05358$+$3543 was not visible at the CSO during our run, only the other three sources were observed in lines of CS\,$J$=$7-6$, SO\,$J$=$8_8-7_7$,  SO$_2$\,$J$=$18_{0,18}-17_{1,17}$, and in corresponding isotopic lines of C$^{34}$S, $^{34}$SO, $^{34}$SO$_2$. The receiver is a helium-cooled SIS mixer operating in double-sideband mode designed to fully cover the 280-420~GHz window. The backend consists of a 1024 channel acousto-optical spectrometer covering a bandwidth of 50~MHz, providing a velocity resolution of $\Delta \varv \simeq 0.086$~\kms\ at 345~GHz. Observations were performed under good but unstable weather conditions ($\tau_{0}^{atm}\sim 0.06$-0.15 at 225 GHz, see Table~\ref{table2} for the T$_\mathrm{sys}$). The pointing accuracy was of the order of 4-5". The CS $J$=$7-6$ line was blended with an H$_2$CO line from the image sideband, therefore we applied a frequency shift to the local oscillator frequency to separate the two lines, but the procedure did not work well for IRAS18264$-$1152 and W43MM1 (hence the CS spectra are contaminated).
	
Data reduction was performed using the CLASS software from the GILDAS suite \citep{guilloteau2000}. We found and eliminated unusable data, subtracted a baseline, then spectra at the same position were summed and finally antenna temperature $T_a^*$ was converted into $T_\mathrm{mb}$ (using the $\eta_\mathrm{mb}$ values from Tab.~\ref{table2}). Special attention was paid to the frequency switching observing method.

%______________________________________________________________

   \begin{figure}
   \centering
   \includegraphics[width=\columnwidth]{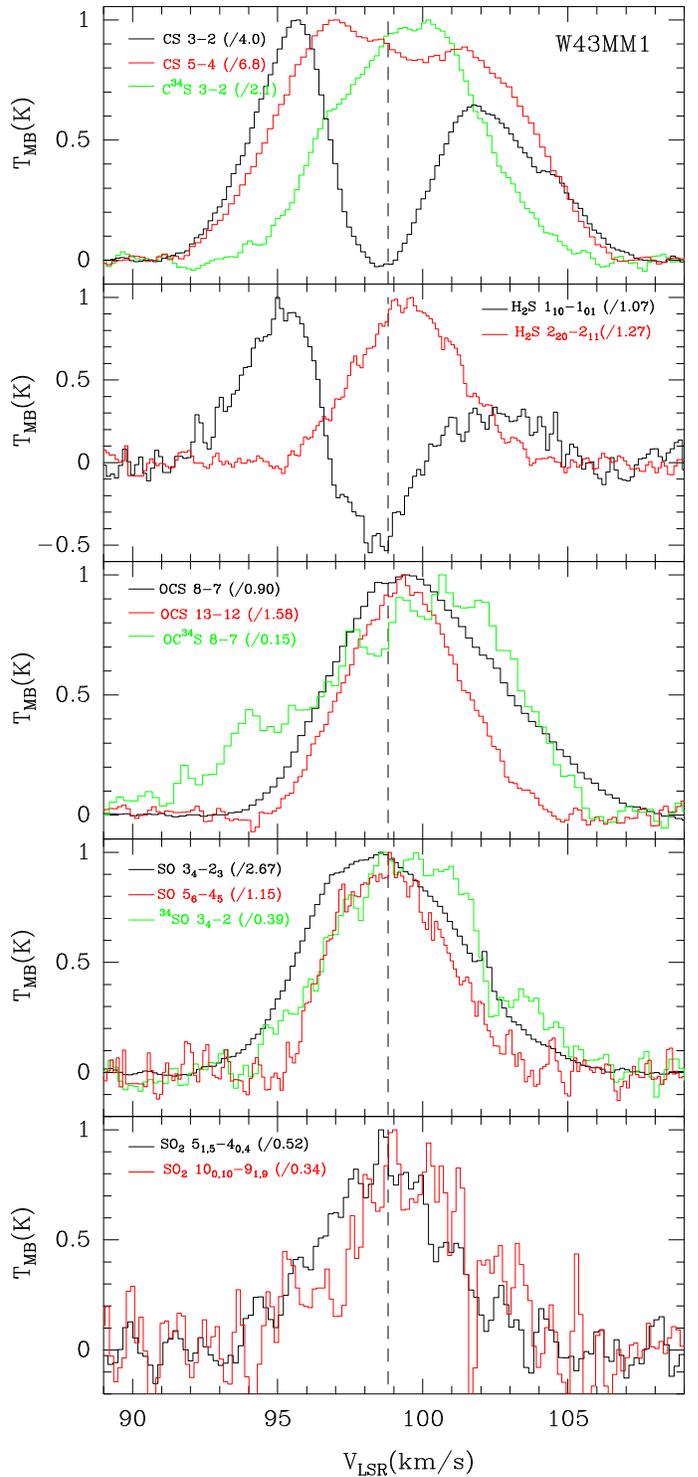}
   \caption{Normalized emission of CS, H$_2$S, OCS, SO, SO$_2$ (and isotopic species) lines from W43MM1. Spectra velocity resolutions are 0.10-0.19~\kms. The dashed line shows the source LSR velocity.}
              \label{FigW43mm1}%
    \end{figure}

   \begin{figure}
   \centering
   \includegraphics[width=\columnwidth]{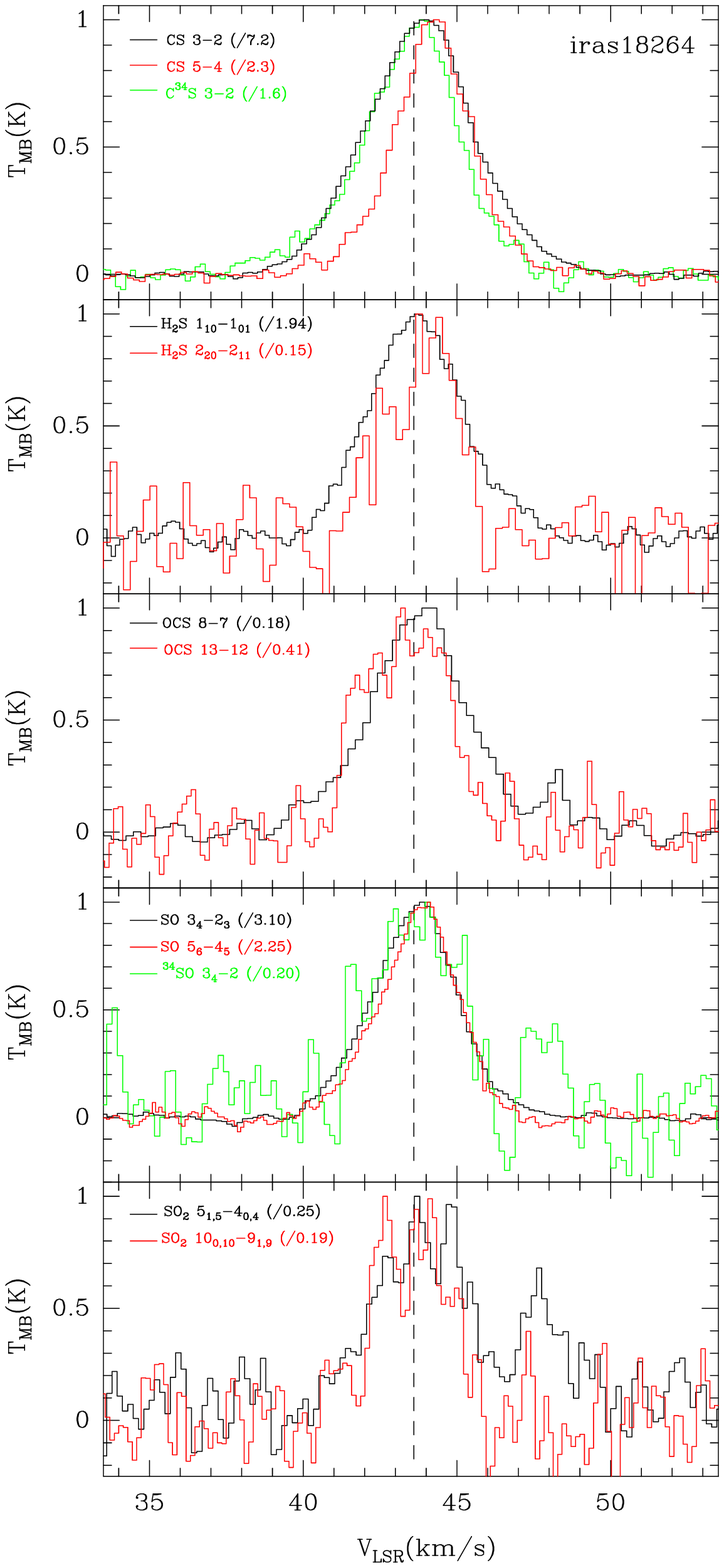}
   \caption{Normalized emission of CS, H$_2$S, OCS, SO, SO$_2$ (and isotopic species) lines from IRAS18264$-$1152. Spectra velocity resolutions are 0.10-0.19~\kms. The dashed line shows the source LSR velocity.}
              \label{Fig18264}%
    \end{figure}

   \begin{figure}
   \centering
   \includegraphics[width=\columnwidth]{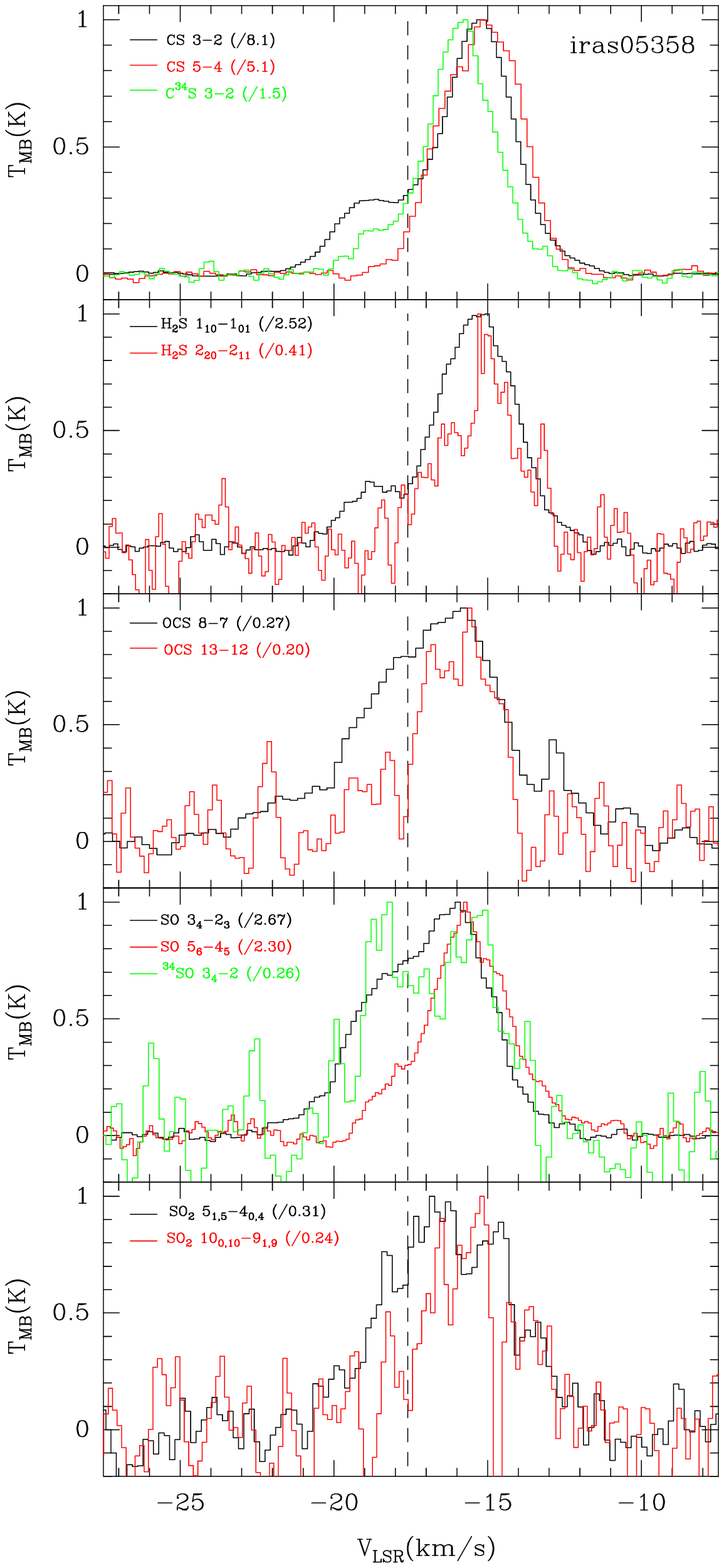}
   \caption{Normalized emission of CS, H$_2$S, OCS, SO, SO$_2$ (and isotopic species) lines from IRAS05358+3543. Spectra velocity resolutions are 0.10-0.19~\kms. The dashed line shows the source LSR velocity.}
              \label{Fig05358}%
    \end{figure}

   \begin{figure}
   \centering
   \includegraphics[width=\columnwidth]{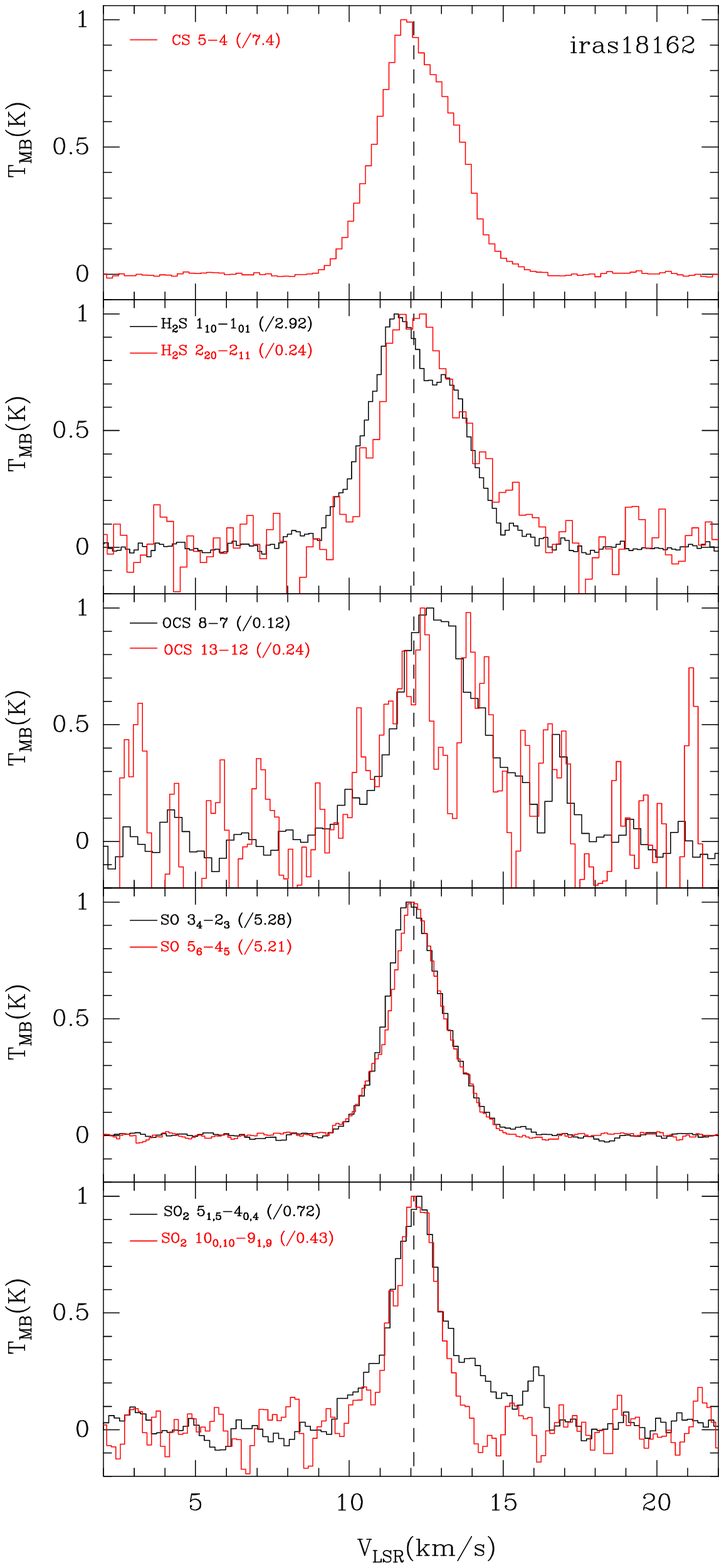}
   \caption{Normalized emission of CS, H$_2$S, OCS, SO, SO$_2$ (and isotopic species) lines from IRAS18162$-$2048. Spectra velocity resolutions are 0.10-0.19~\kms. The dashed line shows the source LSR velocity.}
              \label{Fig18162}%
    \end{figure}

\section{Results}

We present here the result of the observations for each source. Some spectra of the detected lines are displayed in Fig.~\ref{FigW43mm1}--\ref{Fig18162} (the remaining spectra are available as online material). Some transitions ($^{34}$SO\,$8_8-7_7$, SO$_2$\,$28_{4,24}-28_{3,25}$ and $^{34}$SO$_2$\,$5_{1,5}-4_{0,4}$) are not detected in any of our sources. The line parameters are given in Table~\ref{line-W43MM1}--\ref{line-iras18162} (only available in electronic form). For all observations, contamination from lines in the image sideband of the receiver has been checked.

\subsection{W43MM1}
\label{lineW43}

The CS line emission from this source is characterized by what looks like a strong self-absorption (at $\sim -0.2$~\kms   \ from the source velocity) with a stronger blue component compared to the red one. Indeed the self-absorption in the $3-2$ line goes below zero kelvins, splitting the line profile into two well separated components. The 2mm continuum flux is actually high (0.8 Jy), which converts into a brightness temperature of 5.5 K, and can therefore explain the observed dip. Similar, but weaker, self-absorption is observed in the $5-4$ and $7-6$ transitions. However, the isotopic lines are only weakly asymmetric, redshifted with respect to the systemic velocity. The derived opacities (from our model, see Sect. \ref{molmod}) are respectively 0.4 and 0.6 for the C$^{34}$S\,$3-2$ and $7-6$ line emissions, indicating optically thick lines. All lines are very broad, up to 15~\kms. Note that HCO$^+$ and  H$^{13}$CO$^+$\,$3-2$ emission lines observed by \citet{motte2003} have exactly the same profiles as the CS and C$^{34}$S emission lines, respectively. The self-absorption is present in the H$_2$S line emissions too (the isotopic line is not detected), even producing for the $1_{1,0}-1_{0,1}$ line a negative dip, below the continuum level, at the same velocity as CS. The $2_{2,0}-2_{1,1}$  line emission has a different profile  with more redshifted emission.

The SO\,$3_4-2_3$, $5_6-4_5$ and the isotopic $3_4-2_3$ line profiles are also weakly asymmetric, with a more pronounced blueshifted emission. The $6_5-5_4$ and $8_8-7_7$ lines exhibit, on the contrary, a stronger redshifted component.
The SO$_2$\,$5_{1,5}-4_{0,4}$ line is also blue-asymmetric, but all other SO$_2$ lines are more or less symmetrical and gaussian.

All OCS lines exhibit a gaussian profile, slightly red-shifted by $\sim
0.5$~\kms\ from the source velocity, with a red wing. The
  isotopic OC$^{34}$S line profile is complicated with several
  components in emission at 91.8, 94.0 and 95.5 \kms \ (see
  Fig.~\ref{FigW43mm1}). No identification with known species was
  possible (we also checked the frequencies in the image band) and
  instrumental effects have been ruled out. This weak component is not seen in the corresponding OCS line profile. Since W43MM1 is a very
  rich molecular source (see Sect. \ref{W43MM1case}), this could well be real lines.

\subsection{IRAS18264$-$1152}

All line emission profiles are well peaked on the source velocity and
have more or less the same shape, with a blue wing. Nonetheless, a distinct weak emission at 48--50 \kms \ is observed for all transitions of all species. These emissions cannot be due to a contamination from any OFF position as observations have been performed in frequency-switch mode. None of these features correspond to a known molecular line. We propose that this secondary emission comes from a different location than the main one, either from another source within the beam, or from a knot in the outflow. Actually, an SiO outflow component  at 48 \kms, offset by 20$^{\prime\prime}$ has been observed by \citet{qiu2007}. 

All CS lines are detected, except the $7-6$ transition that is blended
with an H$_2$CO line from the image band, hence completely contaminating
the profile. The C$^{34}$S\,$7-6$ line is tentatively detected. The
H$_2$S lines exhibit a less peaked profile with a blue wing too. All
the SO lines are detected too, but the $6_5-5_4$ line emission
exhibits a strong and very broad red wing which looks suspicious. We
do not consider this line for further analysis. The SO$_2$ line
profiles are more flattened. Actually, an unidentified line at
47.8~\kms\ is observed in the SO$_2$\,$5_{1,5}-4_{0,4}$ spectra, and
could come from the reference OFF spectra, just as the suspicious $6_5-5_4$
wing. The OCS lines exhibit similar profiles, with a blue asymmetry.

\subsection{IRAS05358$+$3543}

The CS line emission is strong in both observed transitions (the $3-2$
line emission being the strongest and broadest one) with a strong dip
at the source velocity, even for the isotopic line. The isotopic line
emission here is optically thin (around 0.1 according to our model),
therefore this dip is likely owing to two velocity components rather than to self-absorption. All profiles are asymmetrical with a stronger red component, the blue one being almost completely extinguished (nearly in absorption) for the CS\,$5-4$ line emission. The peak-to-peak velocity separation is 3.0, 3.5 and 4.3~\kms, respectively for the C$^{34}$S\,$3-2$, CS\,$3-2$ and $5-4$ lines. 

The H$_2$S\,$1_{1,0}-1_{0,1}$ line emission also exhibits a strong dip at the source velocity. The separation between the two emission peaks is 3.6~\kms. The same shape is also seen for the $2_{2,0}-2_{1,1}$ line, where no blue peak is seen, the red component being split in two peaks. The isotopic line might be tentatively detected, with a very strong asymmetry. 

Profiles of the SO lines generally exhibit a dip at the source velocity. The peak-to-peak separation is 3~\kms\ for the $3_4-2_3$ line. The isotopic $3_4-2_3$ line shows a red-shifted (by more than 1~\kms) double-horn profile. $^{34}$SO $3_4-2_3$ line emission is marginally detected, suffering from obvious baseline problems. We checked, without success, if the double-peak profile could be due to other species.

The SO$_2$ emission is detected in all transitions (except the $11_{1,11}-10_{0,10}$), but the $S/N$ is quite low (5-9). All the lines are asymmetric, nearly without any red component. The $14_{0,14}-13_{1,13}$ emission is the strongest. 

A dip is seen in all OCS lines too, most prominently in the  $13-12$ transition. The  $19-18$ line is the strongest. The OCS 13--12 line is only tentatively detected with a complex profile.

A second emission component is seen in the C$^{34}$S 3--2, OCS 8--7,
SO $3_4-2_3$, SO$_2$ $5_{1,5}-4_{0,4}$ spectra at --13 \kms. As this emission is also present in the wing of the CS
3--2 and H$_2$S 1--1 line emission (observed with a smaller telescope
beam), we believe that this emission originates from the source mm1b as observed in the C$^{34}$S 5--4 line emission by \citet{leurini2007} at the same velocity. 

\subsection{IRAS18162$-$2048}
 
The CS line shapes are similar for the different transitions observed, with the $5-4$ line being substantially stronger and broader, exhibiting a red wing. 
 
The H$_2$S emission is slightly broader than CS, with more complex profiles: the $1_{1,0}-1_{0,1}$ emission profile is made of two peaks, possibly revealing self-absorption at a redshifted velocity (by 0.5~\kms \ with respect to the source systemic velocity). The $2_{2,0}-2_{1,1}$ line profile is slightly asymmetrical, with a red wing. The optically thin isotopic line emission exhibits a symmetrical profile centered on the source velocity, with a narrower profile than for the H$_2$S emission.

The SO lines are strong and easily detected, but the isotopic line is not detected. The SO$_2$ lines are well detected, even the $18_{0,18}-17_{1,17}$ transition, the $11_{1,11}-10_{0,10}$ being the strongest one. The line profiles show a red wing. 

All the OCS lines are completely asymmetrical with a very weak blue component, indicating a self-absorption at the source velocity.
 
For that source too, emission a few \kms \ away from the main emission is detected (at 16--17 \kms) for the lines observed with a beam larger than 18$^{\prime\prime}$, likely revealing the presence of a second source within the beam.

\section{Analysis}

\subsection{Line Asymmetries}

Analysis of the line profiles can reveal the presence of outflows, infall or even rotation because of their known signatures \citep[see][and references therein]{fuller2005}: outflow or rotation give rise to both red and blue asymmetric lines, while infall produces only blue ones. Indeed, the profiles of optically thick lines from infalling material have stronger blueshifted emission than redshifted one. However, it is important to stress that outflow or rotation could also produce a blue asymmetric line profile along a particular line of sight to a source.

 To quantify the asymmetry of a line, we use \citet{mardones1997} criteria: we calculate the asymmetry parameter
\begin{equation}
\delta \varv=\frac{\varv_\mathrm{thick}-\varv_\mathrm{thin}}{\Delta v_{1/2}}
\end{equation}
where $\varv_\mathrm{thick}$ and $\varv_\mathrm{thin}$ are the line peak velocity respectively of an optically thick and thin tracer, and $\Delta v_{1/2}$ the line width (FWHM) of the optically thin line.

Because not all the same lines are detected or observed in all
sources, the difficulty is to find a common strong optically thin line, generally from isotopic species. We use $^{34}$SO\,$3_4-2_3$ for IRAS05358$+$3543 ($\varv_\mathrm{thin}=-17.0$~\kms, $\Delta\varv_{1/2}=4.4$~\kms), IRAS18264$-$1152  ($\varv_\mathrm{thin}=43.6$~\kms, $\Delta \varv_{1/2}=4.0$~\kms), and W43MM1 ($\varv_\mathrm{thin}=99.2$~\kms, $\Delta\varv_{1/2}=5.5$~\kms), and H$_2$$^{34}$S\,$1_{1,0}-1_{0,1}$ for IRAS18162$-$2048 ($\varv_\mathrm{thin}=12.1$~\kms, $\Delta\varv_{1/2}=1.5$~\kms).  Results are given in Table~\ref{asym}.

\begin{table}[]
 \begin{minipage}[t]{\columnwidth}
 \caption{\label{asym} Asymmetry parameter \citep[][]{mardones1997} calculated for the detected line emissions. Negative values of $\delta\varv$ correspond to blueshifted emission while positive values are redshifted.}\label{asymmetry}
\begin{center}
\begin{tabular}{lccccc} \hline \hline
{\bf Species} & Line\footnote{{\em Reference} lines are optically thin $^{34}$SO and H$_2$$^{34}$S line emissions} &  \multicolumn{4}{c}{Source} \\ 
 & & W43MM1 & 18264 & 05358  & 18162 \\ \hline
CS & 3-2 &  -0.69	& 0.05 & 0.37 &  \\
 & 5-4 & -0.41	& 0.16	& 0.42	& 0.03  \\
 & 7-6 &  -0.40 & \footnote{doubtful line} &	& -0.01 \\ \hline
C$^{34}$S &  3-2 & 0.05 & -0.02  & 0.27 &   \\
 & 7-6  &  0.18 & 0.20 & & 0.13  \\ \hline
SO & $3_4-2_3$ & -0.10 & 0.00 & 0.22 & 0.03   \\
 & $5_6-4_5$ &  -0.06 & 0.05 & 0.29 & 0.06  \\
 &$6_5-5_4$ & 0.11 & \footnote{doubtful line} & 0.28 & -0.08  \\
 & $8_8-7_7$ & 0.11 & 0.04 &  & 0.12 \\ \hline
SO$_2$ 	& $5_{1,5}-4_{0,4}$ &  -0.11 & 0.08 & 0.17 & 0.03 \\
 			& $10_{0,10}-9_{1,9}$ &  0.05 & -0.02 & 0.27 & 0.05 \\
 			& $11_{1,11}-10_{0,10}$ &  & -0.03 & & 0.44 \\
 	& $14_{0,14}-13_{1,13}$ &  &  -0.12 & 0.38 & 0.00 \\
 			& $18_{0,18}-17_{1,17}$ & 0.10 &  & & 0.19    \\ \hline
OCS 			& $8-7$  & 0.11 & 0.06 & 0.26 & 0.53  \\
 			& $13-12$ &  0.03 & -0.06 & 0.24 & 0.52 \\
 			& $19-18$ &  0.10 & 0.09 & 0.17 & 1.08 \\  \hline
H$_2$S 	& $1_{1,0}-1_{0,1}$ &  -0.76 & 0.02 & 0.39 & -0.37    \\
 	& $2_{2,0}-2_{1,1}$  &  0.06 & 0.08 & 0.42 & 0.16 \\ \hline
\hline 
\end{tabular}
\end{center}
\end{minipage}
\end{table}
 
Negative values of $\delta\varv$ correspond to blueshifted emission while positive values to redshifted one. Following \citet{mardones1997} and \citet{fuller2005}, we adopt a criteria of $|\delta \varv| >0.25$ to indicate that a line profile is asymmetric. Main conclusions are that the source IRAS18264$-$1152 has quite symmetric lines, except for the C$^{34}$S\,$7-6$ line (close to the 0.25 value), while for IRAS18162$-$2048 asymmetric lines are observed in OCS, in the H$_2$S\,$1_{1,0}-1_{0,1}$  (redshifted) and in the SO$_2$ 11-10 transitions (blueshifted). Very strong asymmetries are observed for the two other sources: most of the lines from IRAS05358+3543 are strongly redshifted ($\delta \varv$ up to $0.42$); W43MM1 shows clearly blueshifted lines in CS (up to $-0.69$) and H$_2$S (up to $-0.76$), probably indicating the presence of infall towards this source. Nevertheless, we stress again that such an emission profile can be sometimes produced by an outflow too, \textit{e.g.} CO\,$1-0$ line profile towards IRAS18162$-$2048 by \citet{benedettini2004}. Hence, this option cannot be eliminated without further investigations.

\subsection{Possible Infall in W43MM1}
\label{infall}

Assuming the observed CS profile towards W43MM1 is due to infall, we
estimate the infall velocity using the approximate method described in \citet{myers1996} for the optically thick lines with a strong blue asymmetry:
\begin{equation}
\label{ }
v_\mathrm{in} \approx \frac{\sigma^2}{\varv_\mathrm{red}-\varv_\mathrm{blue}}  \ln{\left (\frac{1+ e T_\mathrm{BD}/T_\mathrm{D}}{1+ e T_\mathrm{RD}/T_\mathrm{D}}\right)}
\end{equation}   
where $T_\mathrm{D}$ is the brightness temperature of the dip (assumed optically thick), $T_\mathrm{BD}$ and $T_\mathrm{RD}$ the height of respectively the blue and red peak above the dip, and $\varv_\mathrm{blue}$ and $\varv_\mathrm{red}$ respectively the velocity of the blue and red peak. The velocity dispersion $\sigma$ is obtained from the FWHM of an optically thin line.

Adopting the $^{34}$SO $3_4-2_3$ line as an optically thin tracer, the velocity dispersion in the circumstellar material is estimated to be 5.5~\kms\ and the source velocity to be 99.2~\kms. The derived infall velocities $v_{in}$ are $2.1$, $1.8$ and $2.9$~\kms, calculated respectively for the CS\,$3-2$, $5-4$ and H$_2$S\,$1_{1,0}-1_{0,1}$ line emissions. Using the following formula (where $m$ is the mean mass of the molecule and $r_{out}$ the inner radius of the most outer layer):
\begin{equation}
\label{infallrate}
\frac{dM}{dt}=4\pi {r_{out}}^2 m n v_{in}
\end{equation}
the inferred {\em kinematic} mass infall rate (for a size $r_{out}$ of 3.4~$10^{17}$~cm and a density $n$ of 2.5~$10^6$~cm$^{-3}$, cf. Table 4) is 5.8--9.4~10$^{-2}$~M$_{\odot}/$yr.

\subsection{Outflows in the other sources or multiplicity effect ?}

  \begin{figure}
   \centering
   \includegraphics[width=\columnwidth]{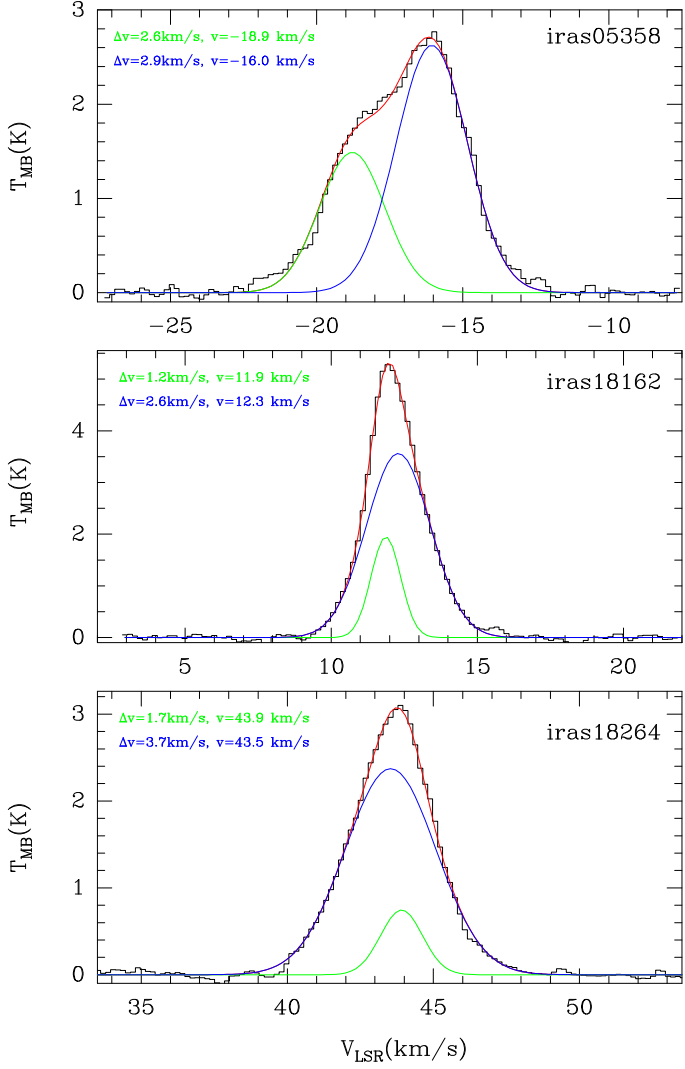}
   \caption{Gaussian fit of the SO\,$3_4-2_3$ line emission from IRAS05358+3543, IRAS18162$-$2048 and IRAS18264$-$1152. The observed spectra are in black, total fit in red, individual components in blue and green.}
              \label{outflow}%
    \end{figure}

The SO\,$3_4-2_3$ lines towards IRAS18162$-$2048 can be fitted with
two Gaussians of FWHM line widths of $1.2$ and $2.6$~\kms \ (see Fig.~\ref{outflow}). The CS\,$7-6$ line emission can be fitted by two Gaussians too: one of $1$~\kms\ broad, the other one of $2.8$~\kms. The optically thin H$_2$$^{34}$S line can be fitted by two Gaussians of FHWM $1.2$ and $3.2$~\kms, comparable to the components derived from the other lines. Moreover, it is interesting to note that the two gaussian components are centered at velocities ($11.9$ and $12.3$~\kms) comparable to the velocities ($11.9$ and $12.8$~\kms) of the two major outflow components seen in CO, $^{13}$CO and C$^{18}$O\,$1-0$ line emissions by \citet{benedettini2004}. For IRAS181264, the CS\,$3-2$ and SO\,$3_4-2_3$ line emissions can be fitted with two Gaussians of two different FWHM: $1.7$ and $3.7$~\kms. These two-components reveal the presence of gas flows, possibly an outflow towards objects IRAS18162$-$2048 and IRAS18264$-$1152. 

Line emission profiles seen in the source IRAS05358+3543 exhibit a strong dip (especially for CS), even for the optically thin isotopic species, which is hence not a self-absorption but rather a the signature of two well separated velocity components with FWHM around $2.6$ and $2.9$~\kms \ (derived from CS\,$3-2$ and SO\,$3_4-2_3$ lines emission fitting, see Fig.~\ref{outflow}). This could be due to the outflow detected by  \citet{beuther2002c} and \citet{beuther2002b} or to the multiplicity of the source itself. IRAS05358+3543 is indeed composed of three mm sources \citep[][]{leurini2007}, all within our beam, mm1 being the most massive, the other two being less evolved (mm3 is probably a starless massive core). Actually, the component at --16 \kms\ is at the same velocity as the source mm2 \citep[][]{leurini2007}. Hence the profile may result from the source multiplicity.

\section{Modelling the continuum and the molecular emission of the sources}

%\subsection{Brief description of the model}

We use the modeling method described in \citet{marseille2008} and
originally developed by \citet{hogerheijde2000}. It consists, in a first step, of constraining the physical structure of the source using a modeling of the SED, and in a second step of applying the obtained physical model to derive abundances of the observed species. In order to reduce the number of free parameters, we restrict our modeling to a simple 1D, radial description of the sources.
The assumed spherical symmetry of the sources is acceptable if we restrict our study to the molecular line emission modeling coming from cold gas ($T <100$~K) as \citet{marseille2008} showed it for the massive dense core case.

  \begin{figure}
   \centering
   \includegraphics[width=\columnwidth]{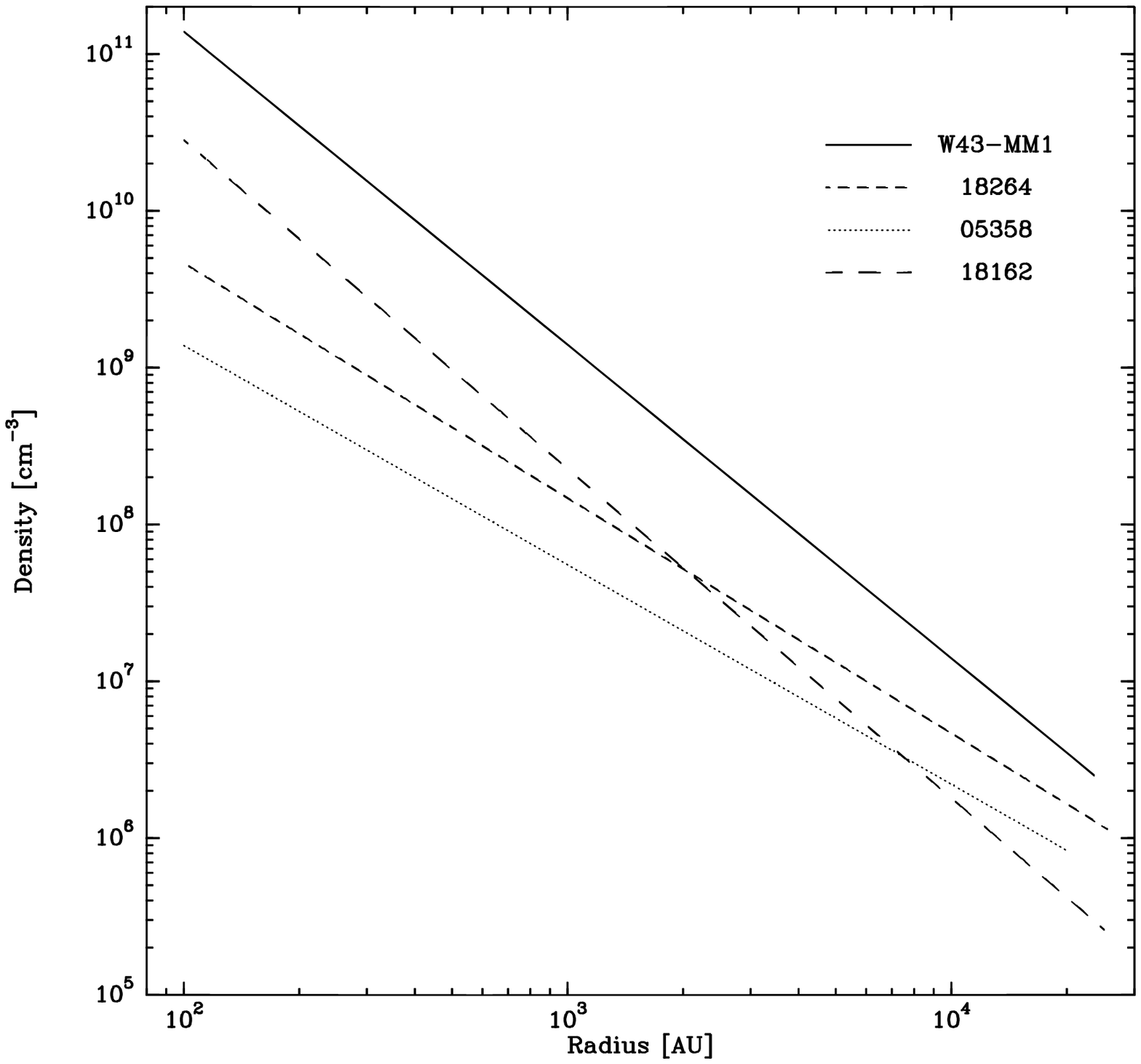}
   \includegraphics[width=\columnwidth]{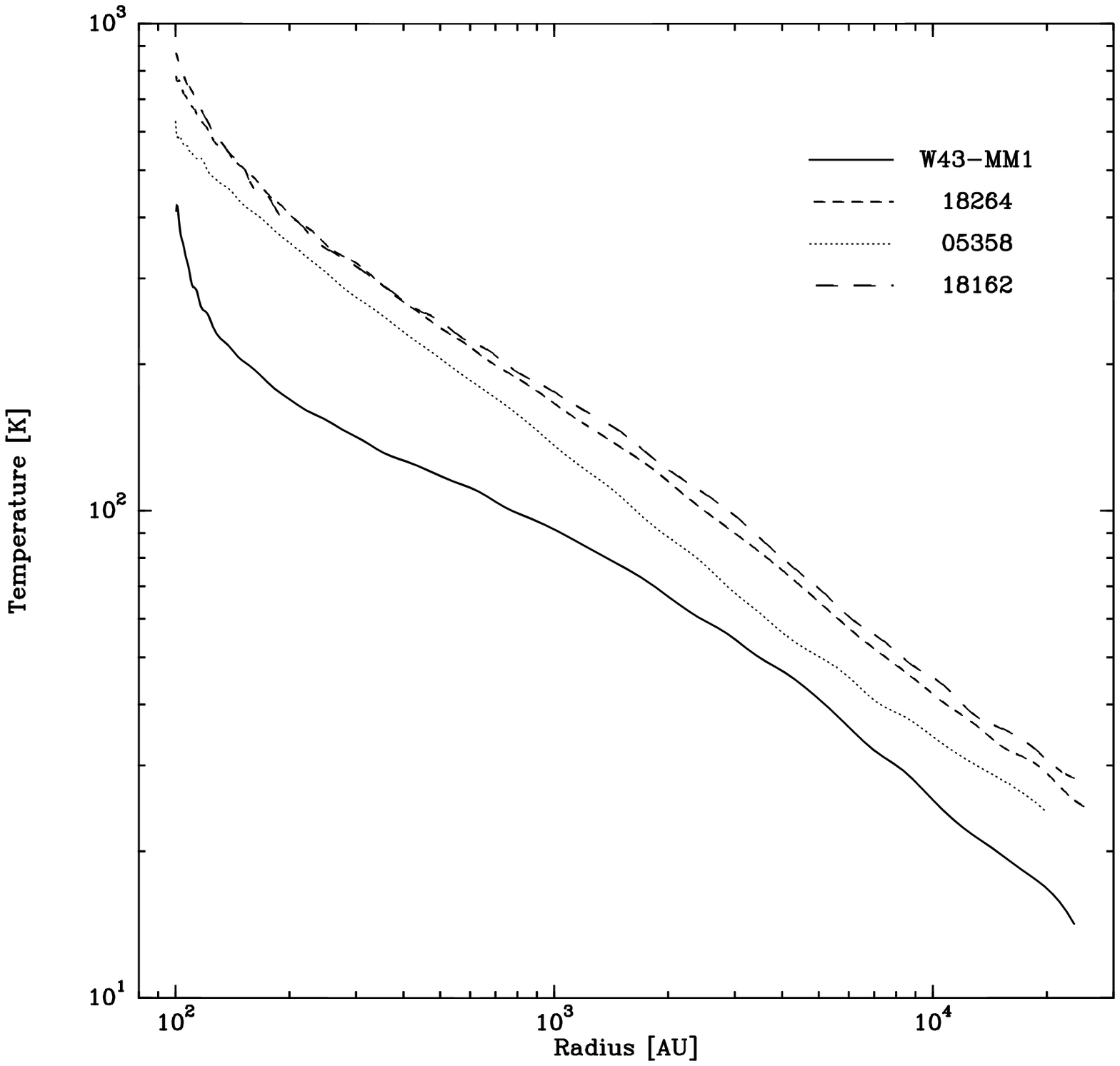}
   \caption{Plot of the density (top) and temperature (bottom) versus the distance to the center for the 4 objects. }
              \label{laws}%
    \end{figure}

  \begin{figure}
   \centering
   \includegraphics[width=\columnwidth]{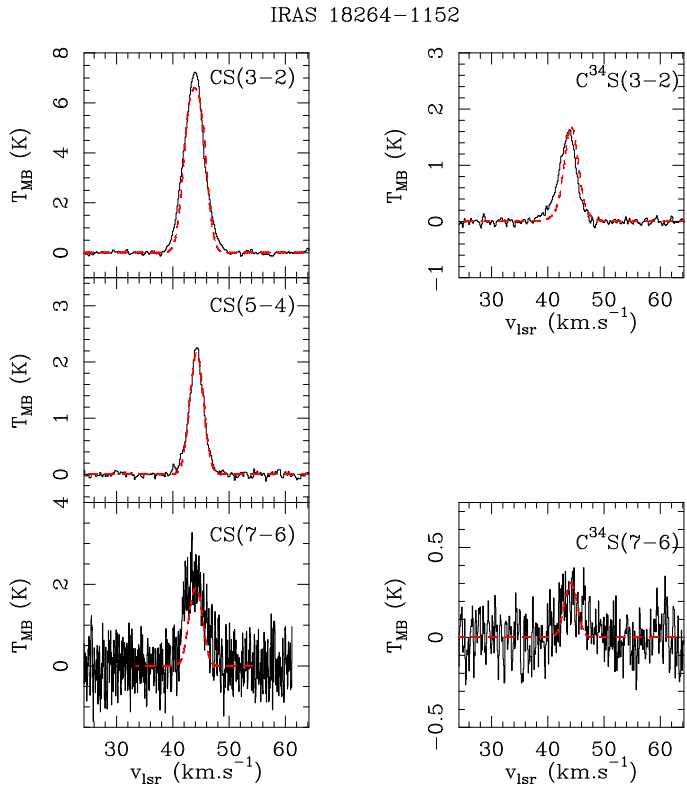}
   \caption{Fit of the CS lines emission from IRAS18264$-$1152. The observed spectra are in black, fit in red.}
              \label{fitCS}%
    \end{figure}
    
\subsection{SED distribution}
\label{seddet}

We use the radiative transfer code MC3D \citep[][]{wolf1999} in its 1D version to model the SED. The model parameters are taken from literature (distance, $d$, power law index, $p$, for density distribution, heating source temperature, $T_*$, and source sizes, $r_{\mathrm{out}}$, the total luminosity, $L$). The parameter, $p$, which defines the density distribution through a power law in the form $\rho(r) \propto r^{p}$, was derived by \citet{beuther2002b} and \citet{motte2003} from fits to the millimeter continuum emission maps. Concerning the source size, we adopt systematically the size of the continuum mm-wave emission, deconvolved by the beam.  These values are summarized in Table \ref{sourcemodel}. The actual value used for the inner radius of the physical structure has virtually no observable effect on the results, we therefore adopt for the inner radius ($r_{\mathrm{sub}}$) the radius of dust sublimation taken at 1500~K.

The fit of the millimeter and submillimeter SED of each source allows deriving the best temperature and distributions by iterating on the total mass which is controlled by the value of the density $n_0$ at reference radius $r_0$ taken equal to 100~AU, and on the average temperature. The values obtained, with mean densities and temperatures of the sources, are reported in Table \ref{sourcemodel} and shown in Fig.~\ref{laws}.

\begin{table}
\begin{minipage}[t]{\columnwidth}
 \caption{Summary of results from dust continuum emission modeling (above mid-line) and from molecular line emission modeling (below mid-line). Different turbulent ($\varv_\mathrm{T}$) or infall ($\varv_\mathrm{infall}$) velocities are used to reproduce the different line emissions from each source.}\label{sourcemodel}
\begin{center}
	\begin{tabular}{c|cccc} \hline \hline
	Source & W43MM1\footnote{source description from Motte et al. 2003} & 18264 & 05358 & 18162  \\ \hline
	$d$ (kpc)								& 5.5	& 3.5	& 1.8	& 1.9 \\
	$L$ ($10^4 \textrm{L}_{\odot}$)	& 2.3	& 1.4	& 0.7	& 2.9 \\
	$T_*$ ($10^4$~K)							& 3.0	& 2.9	& 2.6	& 3.0 \\
	$r_{\textrm{out}}$ ($10^4$~AU)			& 2.5	& 2.7	& 2.1	& 2.6 \\	
	$p$\footnote{from Beuther et al. 2002, excepted for W43MM1}	& -2.0	& -1.5	& -1.4	& -2.1 \\
	\hline
	$n_0$\footnote{density at 100~AU}($10^4$)	& 40	& 1.0	& 0.3	& 6.1 \\
	$r_{\textrm{sub}}$ (AU)					& 4.9	& 26.5	& 20.1	& 27.2 \\
	$\alpha$\footnote{$Log(T) = \alpha \ Log(r) + \beta$}						& -0.54	& -0.61	& -0.62	& -0.60\\
	$\beta$									& 3.55	& 4.05	& 3.99	& 4.04 \\
	$\left<T\right>$	(K)						& 20.0	& 32.6	& 31.1	& 35.6 \\
	$T_{\textrm{out}}$ (K)					& 14.2	& 24.6	& 24.1	& 27.2 \\
	$\left<n\right>$ ($10^6$ cm$^{-3}$)		& 11.8	& 2.6	& 0.2	& 1 \\
	$M$ (M$_{\odot}$)						& 4100	& 1200	& 400	& 570 \\
	$v_t$ (km$/$s) & 0.8-1.5 & 0.7-1.3 & 0.7-2.6 & 0.5-1.8 \\
	$v_{infall}$ (km$/$s) & -(3.5-0.5) & & & \\
	\hline
	\hline
	\end{tabular}
\end{center}
\end{minipage}
\end{table}

   \begin{figure}
   \centering
   \includegraphics[width=\columnwidth]{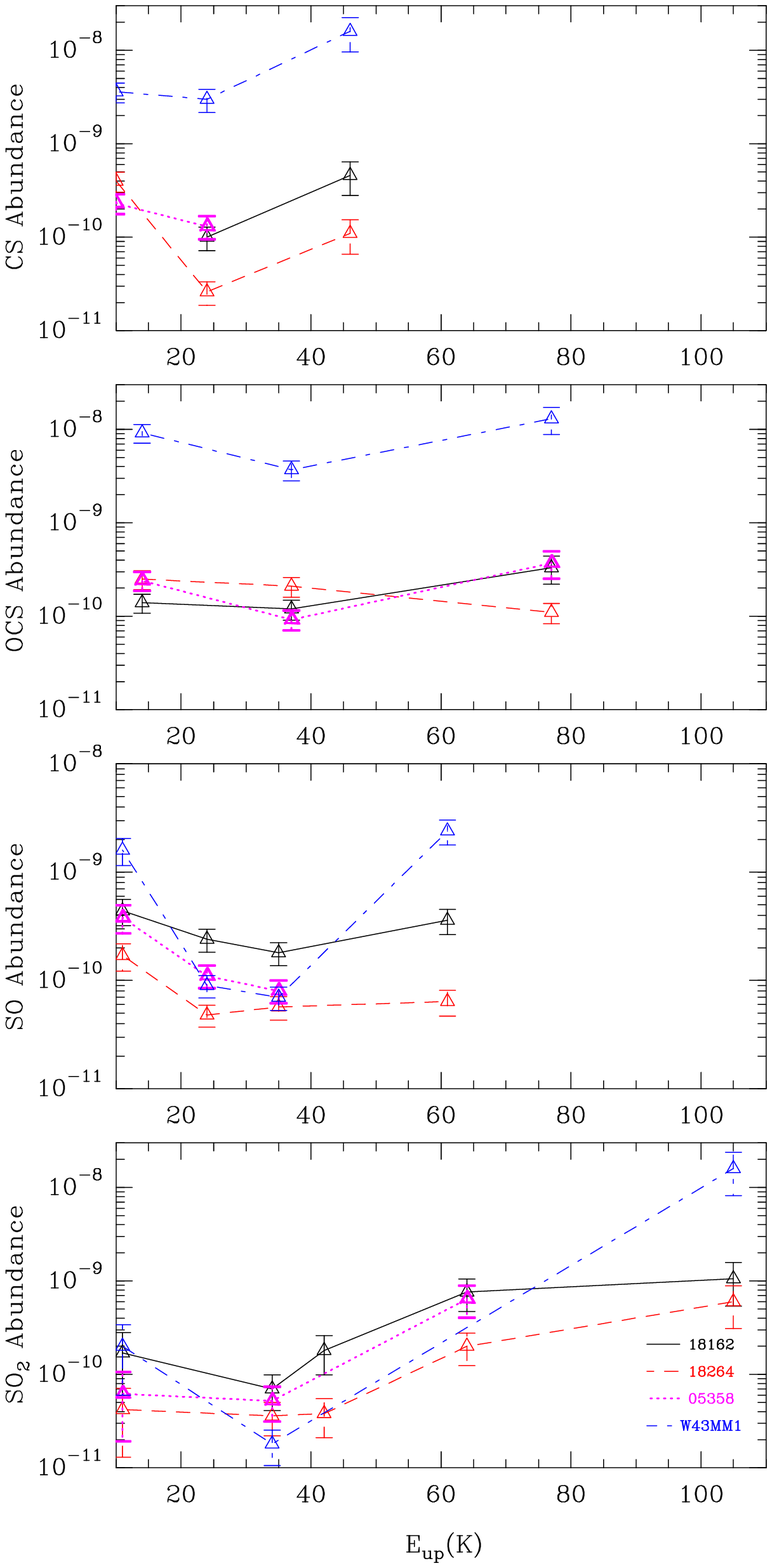}
   \caption{Abundances of CS, OCS, SO and SO$_2$ (relative to H$_2$) versus the energy of the upper level of the transition given in Table \ref{abundances} for the 4 sources. }
              \label{FigEup}%
    \end{figure}

\subsection{Modeling of the molecular emission}
\label{molmod}

The physical description of the sources obtained from the SED modelling is then used to model the molecular emission. We use the RATRAN radiative transfer code developed by \citet{hogerheijde2000}. We build data cubes with a velocity resolution equivalent to the observations, and with a high spatial resolution (0.5\arcsec) which are convolved with the beam sizes of the observations for a direct comparison with the observed spectra. We adopt constant abundances for the RATRAN modeling. The abundances of all species are derived independently for each observed transitions of each species. Then the different abundances obtained for each particular species are interpreted as due to change in abundance as a function of the excitation temperature and therefore as a function of depth into the sources.

The line profiles observed are also affected by gas motions, particularly through the turbulence which is mainly responsible for the line emission widths seen in massive protostars. Thus we add a turbulent velocity field characterized by a rms velocity $\varv_\mathrm{T}$, for each transition of each molecular species (actually OCS lines are the most sensitive to the gas components in our model). In the case of W43MM1, where asymmetric profiles possibly due to infall are detected, we introduce a radial velocity field that follows a law in the form $\varv_{\textrm{inf}} = \varv_0/\sqrt{r/r_{\mathrm{out}}}$, and fit the asymmetry parameter with different values of $\varv_0$ for each transition and molecular species. Concerning the other sources, when the line emission profiles are clearly asymmetric, we are not able to reproduce those line profiles because:
\begin{itemize}
  \item the modelling does not take into account the sub-structure (i.e. multiplicity of the massive dense cores, as for IRAS05358+3543) which is hence not reproduced correctly in the derived source model, 
  \item outflows are not included in the model while they strongly influence the line profiles.
\end{itemize}
Some results for CS in the source IRAS18264$-$1152 are shown in Fig.~\ref{fitCS}.

\begin{table*}
%\begin{minipage}[t]{\columnwidth}
 \caption{Molecular abundances derived for all species within the 4 sources. The temperature of the layer where most of the emission comes from is given. The column 4 gives the line whose emission is the strongest at this distance. Abundances for other sources (from the literature) are given in the columns 9, 10 and 11 for comparison. AFGL2591 is a mid-IR bright HMPO, G31.41+0.31 a hot core, and IRAS16293 a low-mass protostar.}\label{abundances}
 \begin{center}
	\begin{tabular}{cccccccc|ccc} \hline \hline
	 {\bf Source} & &  &  & W43MM1 & 18264 & 05358 & 18162 & AFGL2591$^a$  & G31.41$^b$ & 16293$^c$\\
	\hline
	{\bf Abundance} & $T$ & {\bf $d^d$} & Line &  & & & & & &  \\  
	   ($ \times 10^{-10}$) & (K) & ($10^3$ AU) & & & & &  & & &\\ \hline
	 {\bf $\chi$(CS)} & 60 & 5.9 & $3-2$ & 36 & 4.0 & 2.3	& -- & 100 & 70 & 5100 \\
	   & 75 & 4.9 & $5-4$	& 30 	& 0.26	& 1.3	& 1.0 & & & \\
	 & 96 & 3.3 & $7-6$	& 160		& 1.1	& --	& 4.6 & & &	 \\
	\hline
	 {\bf $\chi$(C$^{34}$S)}& 60 & 5.9  &$3-2$	& 1.8	& 0.32			&	0.27 	&	--	& & & \\
	& 96 & 3.3 &$7-6$		& 8.0	& 0.50 		&	--			& 	0.27	& & & \\
	\hline
	 {\bf $\chi$(OCS)}& 72 & 5.2 &$8-7$		&	92	&	2.5 	&	2.4 	&	1.4	& 100 & 40 &10000 \\
	& 76 & 4.8 &$13-12$		&	37	&	2.4 	&	0.92 	&	1.2	& & & \\
	 & 96 & 3.3 &$19-18$		&	130	&	1.1	&	3.7 	&	3.3	& & &\\
	\hline
	 {\bf $\chi$(OC$^{34}$S)}& 72 & 5.2 &$8-7$			&	4.6	&	$<$0.07	&	$<$0.35 	&	$<$0.25 & & & \\
	\hline
	{\bf $\chi$(H$_2$S)} & 60 & 5.5 &$1_{10}-1_{01}$	& 3.0 & 1.3	&	1.2	&	3.1	& 80 & $\geq 30$ & 5300  \\
	 & 96 & 3.3 &$2_{20}-2_{11}$ & 4.0 &  0.15 & 0.51 & 0.30 & & \\ 
	\hline
	 {\bf $\chi$(H$_{2}^{34}$S) }& 60 & 5.5 &$1_{10}-1_{01}$	& $<$0.19 	& 0.11	&	$<$0.13		&	0.21 & & & \\
	\hline
	{\bf $\chi$(SO)} & 51 & 8.7 &$3_4-2_3$	& 	16	& 	1.7	&	3.8	&	4.4	& 100 & 40 & 17000  \\
	 & 62 & 6.4 &$5_6-4_5$	&	0.90	&	0.48 	&	1.1	&	2.4	& & & \\
	 & 75 & 4.8 &$6_5-5_4$	&	0.70	&	0.57	&	0.8	&	1.8	& & & \\
	 & 96 & 3.3 &$8_8-7_7$	&	24	&	0.64 	&	--			&	3.6	& & & \\
	\hline
	 {\bf $\chi$($^{34}$SO)}& 51 & 8.7 &$3_4-2_3$	& 	0.80	& 	0.11	&	0.33	&	$<$0.06	& & & 	\\
	  & 96 & 3.3 &$8_8-7_7$	&	$<$4.0		&	$<$1.0	&	--			&	$<$0.5	& & & \\
	\hline
	 {\bf $\chi$(SO$_2$)}& 57 & 7.2 &$5_{1,5}-4_{0,4}$		& 	2.0	& 	0.42	&	0.62	&	1.7	& 20 & 120 & 5400 \\
	 & 71 & 5.3 &$10_{0,10}-9_{1,9}$		&	1.8	&	0.36	&	0.52 	&	0.70	& & & \\
	 & 76 & 4.8 &$11_{1,11}-10_{0,10}$	&	--	&	0.38	&	--			& 1.76	& & & \\
	 & 112 & 2.7 &$14_{0,14}-13_{1,13}$	&	-- 	& 2.0	&	6.4		&   7.6	& & & \\
	 & 112 &  2.7 &$18_{0,18}-17_{1,17}$	&	160	& $<$6	&	--			&	10.6	& & & \\
	\hline
	 {\bf $\chi$($^{34}$SO$_2$)}& 57 & 7.2 &$5_{1,5}-4_{0,4}$		& $<$0.33		& $<$0.44		&	$<$16	&	$<$0.22	& & & \\ \hline
	\end{tabular}
\end{center}
$^a$ Mean abundances from van der Tak et al. 2003 \\
$^b$ Mean abundances from Hatchell et al. 1998 \\
$^c$ Mean abundances from Wakelam et al. 2004a \\
$^d$ indicative corresponding distance to the center for the source IRAS18162$-$2048 \\
%\end{minipage}
\end{table*}

\subsection{Uncertainties}

An overview of uncertainties in the modeling process that we used shows that precision on our abundance results are influenced by four main points: the signal-to-noise ratio, the total mass of the source in the model, its temperature and the populations of energy levels of the molecule studied. Furthermore, an absolute error comes from the uncertainty on the dust opacity at millimeter wavelengths, but it disappears in the context of our work where the use of the same radiative transfer code and the same modeling process keep the comparison between each sources relevant. While signal-to-noise ratio is directly known from observations, the other uncertainties have to be detailed and derived to determine how important they are.

At first, mass uncertainty comes from measurements of the flux at
millimeter wavelengths. Considering calibration errors, we can assume
that this value is known with a 20~\% accuracy. Error on temperature
is directly linked to this first error, in addition with the little
variation of the bolometric luminosity induced. Nevertheless, results
of the radiative transfer models show that extreme cases make the temperature varying by 3~\% at the maximum, and tend to compensate first error in the total mass. Even if temperature does not change a lot, its influence on energy level populations, combined with total mass (hence density) uncertainty, can be strong enough to be significant. These uncertainties are dependent on the molecule studied and the transition observed. As we noted previously, the errors can compensate themselves, especially when variation of the level energy populations of a transition are going the same way.

Finally, we derived rough values of uncertainties by testing
temperature and energy level population variations induced by flux
errors at millimeter range. Results show that our modeling process has
a relative precision between 25~\% and 30~\%, depending on the
considered molecular line emission. This range of values does not take
into account the signal-to-noise ratio of the observations. Concerning
SO$_2$ transitions, populations of the energy levels are varying
widely, making the uncertainty increase up to 40~\%, even 70~\% for
the $5_{1,5}-4_{0,4}$ where populations have an opposite behaviour
against temperature variations in our models. Furthermore, we note
that optically thin emission was assumed, whereas some of them are
clearly thick (see Table~\ref{opacities}). That is why abundances derived in this case must be treated as order of magnitude indicative values.

\subsection{Molecular abundances: results}

The derived abundances for each molecule and each source are given in Table~\ref{abundances} for different layers. Each layer is characterized by a temperature (and of course by a density, but the population of high-energy levels is sensitive to the temperature for dense gas tracers).  We also give the strongest line emission in that layer (but of course lines are excited over several layers). Results are plotted in Fig.~\ref{FigEup} versus the energy of the upper level of the transition. Obviously emission lines coming from warmer regions (more than 120~K), \textit{e.g.} involving higher upper energy levels, are missing to probe the most inner regions.

The H$_2$S abundance is roughly the same in the 4 objects studied ($1.6-3.6$~$10^{-10}$) in the outer region ($T\sim60$~K), showing more variation ($0.16-4$~$10^{-10}$) in the warmer inner parts ($T\sim100$~K): in IRAS18264$-$1152, IRAS05358+3543 and IRAS18162$-$2048, the abundance is one order of magnitude lower while it remains constant in W43MM1.

An obvious trend is seen for SO$_2$, likely because the energy
  ladder of this molecule is best sampled. Its abundance is
  roughly constant (or decreases for W43MM1) in colder layers ($T<80$
  K, \textit{e.g.} $>5$~$10^3$~AU for the source IRAS18162$-$2048, and
  E$_\mathrm{up} <$ 35~K), then increases by one order of magnitude in
  warmer layers (hence outer to more inner parts), \textit{e.g.} $\sim
  3$~$10^3$~AU for the same source. Concerning all the other
  molecules, as fewer transitions have been observed, i.e. the energy
  ladder is not completely covered, no obvious trends are seen. In
  spite of an insufficient energy coverage, the derived OCS abundance
  might weakly follow the same trend as seen for CS, except in the source IRAS18264$-$1152. We note that the sources IRAS18264$-$1152 and IRAS05358+3543 exhibit the same abundance variations for SO and SO$_2$. 

In the three less evolved objects, CS and OCS are the most abundant species in the colder outer region while it is SO for IRAS18162$-$2048 (for that source OCS is less abundant than SO and no CS 3--2 has been observed), SO$_2$ being the least abundant.

Reproducing the whole CS and C$^{34}$S emission is impossible with the same isotopic ratio for all the sources.The derived isotopic ratio [$^{32}$S$/^{34}$S] is $9-12$, $10-15$, $15-20$ and 20 respectively for IRAS05358+3543, IRAS18264$-$1152, IRAS18162$-$2048 and W43MM1. The Solar system abundance ratio is 22.5 \citep[][]{kahane1988}, but \citet{chin1996} found ratios increasing from 14 to 35 with the distance to the Galactic center, but this is in contradiction with \citet{nilsson2000} who derived a ratio of 11 towards the very distant source W49N.

\section{Discussion}

%\subsection{Molecular abundances versus evolutionary status}

\subsection{W43MM1: a special case}

\label{W43MM1case}

W43MM1 is clearly different from the other sources, with the highest abundances (relative to H$_2$) of CS, OCS, SO (except in the flattening region); the SO$_2$ abundance in W43MM1 is the highest in the warmer inner part. Moreover the plots (see Fig. \ref{FigH2S} and \ref{FigH2S100}) of the CS and OCS molecular ratios relative to H$_2$S underline this difference, showing high ratios for that source compared to the others. 

The reason might be that W43MM1 though being a very young source has
already developed a hot core with temperatures larger than 200 K
\citep[e.g. CH$_3$CN lines observed by][]{motte2003}. This hot core
region might also be at the origin of the unidentified lines described
in Sect.~\ref{lineW43}. The presence of this hot core might also
explain the high abundance of H$_2$S and OCS molecules which
evaporation from grain surface could be important
\citep[][]{wakelam2004a}, and consequently the high abundance of SO.
The CS case is different because this molecule is already present in
the gas before the evaporation. Actually the CS chemistry is closely linked
to the atomic carbon abundance which might be higher in W43MM1 than in the other sources.

The very high SO$_2$ abundance close to the center of W43MM1 may be due to strong shocks occurring in the inner part of the envelope. The fact that W43MM1 is a very young source (as indicated by the possibly detected infall, see Sect. \ref{infall}) suggests that this massive dense core harbors one or several high-mass protostars with powerful outflows, which might induce shocked regions. Strong shocks can convert SO into SO$_2$ \citep[see][]{hatchell1998}, leading then to a ratio of unity, as observed in the other sources where the process has already taken place, transforming the ratio all across the envelope. Hence shock chemistry might contribute to the SO$_2$ formation process.

Actually,  IRAS05358+3543 harbors a hot core too
  \citep[]{leurini2007}, but very likely more compact than the one of
  W43MM1 \citep[]{motte2003}, hence more diluted in the beam of our
  observations. Moreover, \citet{leurini2007} suggested that the
  IRAS05358+3543 hot core might not produce a very rich chemistry
  (indeed, these authors underlined differences between its molecular spectrum and the one of typical massive hot cores).

\subsection{Molecular ratios as a diagnostic tool of evolution ?}

Sulphur is released into the gas phase in the form of H$_2$S, which is
subsequently transformed into
SO and later into SO$_2$ via neutral-neutral reaction at $100$~K on
time scales of $\sim 10^3$~yr \citep[][]{wakelam2004a}. The initial
destruction of H$_2$S by H$_3$O$^+$ is even more efficient in
the high-mass objects than in low-mass protostars, as water is more abundant here \citep[see][]{vandertak2006}. At higher temperature (T$\geq 300$ K), H$_2$S can be produced again. Once in the gas phase, OCS is destroyed later than H$_2$S (by S$^+$ and cosmic rays). CS destruction at low temperature produces SO and SO$_2$ too, while CS is produced again at higher temperature from OCS. The time dependence of these process might allow us to see an evolution of the abundance along the protostar evolutionary track. 

   \begin{figure}
   \centering
   \includegraphics[width=180pt, angle=270]{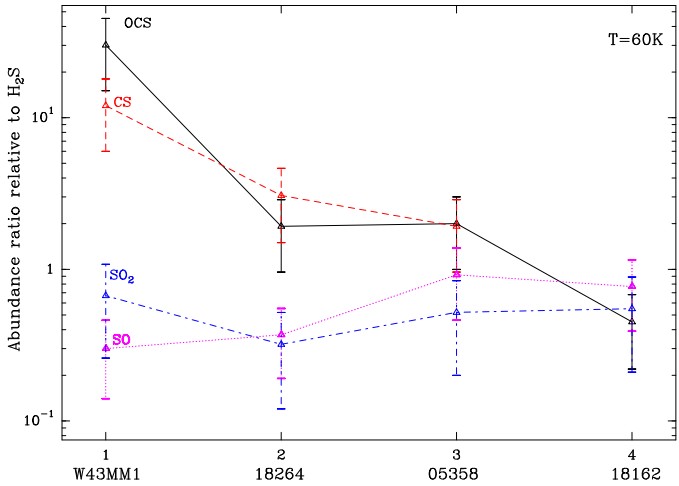}
   \caption{Abundance ratio of OCS, CS, SO and SO$_2$ relative to H$_2$S for each source in the layer at $T=60$~K. The sources are ordered according to the SED classification. }
              \label{FigH2S}%
    \end{figure}

   \begin{figure}
   \centering
   \includegraphics[width=180pt, angle=270]{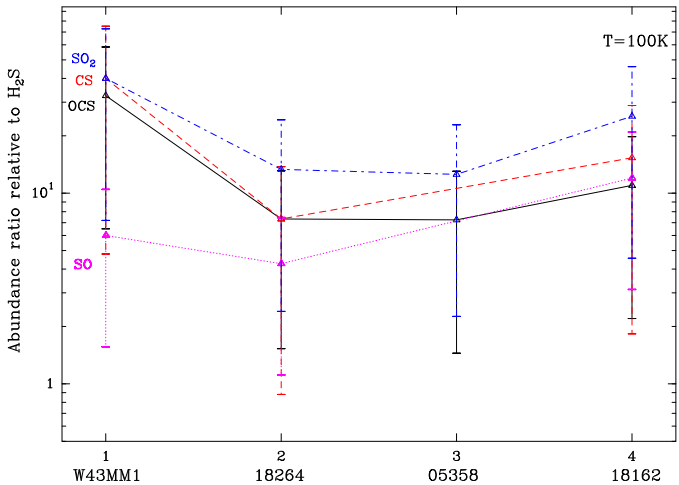}
   \caption{Abundance ratio of OCS, CS, SO and SO$_2$ relative to H$_2$S for each source in the layer at $T=100$~K. The sources are ordered according to the SED classification. }
              \label{FigH2S100}%
    \end{figure}

Actually no evolution is obviously seen for the abundances
  relative to H$_2$, even if they differ significantly from source to
  source. Nevertheless, W43MM1 shows larger abundances of CS, OCS and
  SO compared to IRAS18264$-$1152 and IRAS05358+3543 (see Sect. 6.4). 

\begin{figure}
   \centering
   \includegraphics[width=\columnwidth]{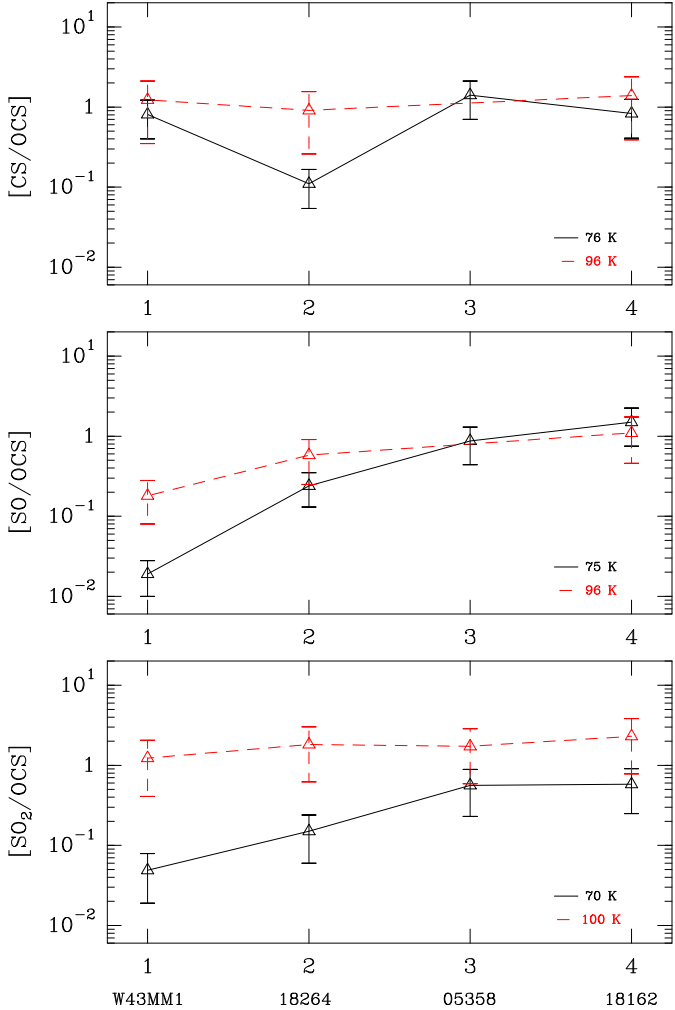}
   \caption{Abundance ratio of CS, SO and SO$_2$ relative to OCS for each source in two different layers. The sources are ordered according to the sED classification. }
              \label{FigOCS}%
    \end{figure}

\begin{figure}
   \centering
   \includegraphics[width=\columnwidth]{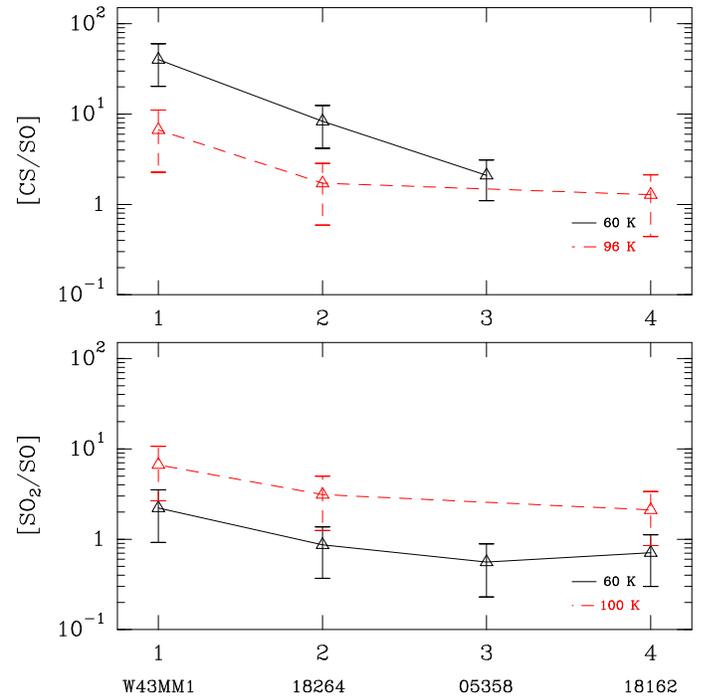}
   \caption{Abundance ratio of CS and SO$_2$ relative to SO for each source  in two different layers. The sources are ordered according to the sED classification. }
              \label{FigSO}%
    \end{figure}

We have investigated possible correlations of the molecular abundance ratios with the order of evolution of our 4 objects for two different layers: one around $T=60-70$~K and a more inner one around $T=100$~K. To be coherent, we only consider values derived for the same layer, i.e same temperature (\textit{e.g.} transitions CS\,$5-4$  and OCS\,$13-12$ with E$_\mathrm{up}=24-37$~K, hence mostly excited in the $75$~K layer). Figures~\ref{FigH2S}, \ref{FigH2S100}, \ref{FigOCS} and \ref{FigSO} show that some global trend appears, either a decrease or an increase with the evolution, but depending on the layer.

We first compare the molecular abundances relative to H$_2$S whose abundance is assumed to remain globally constant during the studied evolutionary period and hence comparable from source to source. In the colder region ($T=60$~K, Fig.~\ref{FigH2S}), the [OCS/H$_2$S] ratio is clearly decreasing from W43MM1 to IRAS18162$-$2048 by a factor 10. This trend is also visible for CS. The [SO/H$_2$S] and [SO$_2$/H$_2$S] ratios remain roughly constant; a small increase might actually be present for SO. In the inner, hence warmer, part of the envelope ($T=100$~K, Fig.~\ref{FigH2S100}), drawing conclusions is more difficult, because not all 4 sources were observed in the same transitions. Moreover, again, W43MM1 stands out as it shows the highest ratio (except for SO), independently of the trend  seen for the three other objects. Actually, according to the temperature and density distributions derived by our model (see Fig.~\ref{laws}), the 100~K layer for W43MM1 is closer to the center of the massive core than in the other sources, and moreover corresponds to a higher density. But the other reason could be that the hot core is dominating. Hence, considering the three other sources, an increase of the different ratios relative to H$_2$S from IRAS181264 to IRAS18162$-$2048 is seen. Comparing the 60 and 100 K regions, the main difference is for OCS and CS: ratios are decreasing at $T=60$~K, but increasing at higher temperature. 

Relative to OCS (Fig. \ref{FigOCS}), the abundance of CS does not show
any trend, neither at 76 K nor at 96 K. At $70-75$~K, the [SO/OCS] and
[SO$_2$/OCS]  ratios increase by at least one order of magnitude along
the derived evolutionary track. At higher temperature, hence in the
inner regions, the same trend is observed for the [SO/OCS] ratio, while [SO$_2$/OCS] remains approximately constant. 

For both probed layers, the [CS/SO] and [SO$_2$/SO] ratios decrease (Fig.~\ref{FigSO}), the variation being more important at 60 K. 

Cross-comparison of the plots for outer regions reveals that the OCS
(and maybe CS) abundance (relative to H$_2$S) decreases while the SO
and SO$_2$ relative abundance (compared to other molecules) increase
along the object evolution. This trend is less obvious at higher
temperature, maybe because in these regions the H$_2$S transformation
process into SO and SO$_2$ \citep[][]{wakelam2004a,wakelam2004b} is
becoming less efficient, as it gets close to the completion: the {\em
  new} SO and SO$_2$ molecules have already been produced in the inner
parts. This hypothesis is strengthened by the larger SO$_2$ abundance
measured in all the objects for the larger temperatures, \textit{i.e.}
in warmer environments. Nevertheless, any conclusions for inner parts
are tricky, because of the incomplete energy ladder coverage of our observations and so the lack of very high energy transitions.

\subsection{Comparison with previous studies}

A comparison between abundances derived for low- and high-mass protostars is not straightforward, because of the different composition of the ices (from which H$_2$S evaporates) and of the different physical conditions (density and temperature; \textit{e.g.} density can be two orders of magnitude higher for the same temperature) as stressed by \citet{wakelam2004a}. Nevertheless, Table \ref {abundances} also gives abundances for AFGL2591 \citep[][]{vandertak2003}, a {\em mid-IR bright} HMPO ($L_{bol}$ $2-10$ times larger, warmer SED with higher $F_{12}$ and $F_{35}/F_\mathrm{total}$), hence more evolved object, and G31.41+0.31 \citep[][]{hatchell1998}, a massive hot core and IRAS16293 \citep[][]{wakelam2004a}. We stress that those values are mean abundances across the whole envelope. Values for AFGL2591 and G31.41 differ a lot for H$_2$S and SO$_2$ (by at least a factor $10$) but are comparable for CS, OCS, and SO to the  abundances found in W43MM1. The other sources from our sample exhibit abundances one or two order of magnitude lower. Abundances in low-mass protostar are $100-1000$ times larger. Clearly the same set of lines has to observed towards these sources and the same modelling applied in order to be able to be compared with our sample.

The sulphur chemistry was studied by \citet{vandertak2003} in the envelopes of more evolved massive protostellar object, {\em mid-IR bright} HMPOs. The abundances they derived are $10-100$ times higher than ours (see Table~\ref{abundances}). They concluded that shock chemistry is unlikely to contribute and that OCS is a major sulphur carrier in the ices. 
%The main bias in their study were that no abundance variation across the envelope was used and that the same physical model was adopted %for all sources. This is critical, as \citet{wakelam2004a} have demonstrated that different local physical conditions even within a single source %(\textit{e.g.} IRAS16293) change the derived abundances. Moreover, van der tak et al. compare absolute abundances, which depend a lot on %the source size and hence do not reveal any vairation from source to source. 
The molecular ratios computed from their work do not show significant variations from source to source. 

%Chemical models for hot cores \citep[e.g.][]{chapman2008} predict that OCS is detroyed after $10^4$ yrs before reaching a constant abundance after $10^6$ yrs. The CS abundance first increases to a peak near $10^4$ yrs, then decreases a little before remaining constant. According to \citet{chapman2008} model for a hot core, [CS/OCS] ratio increases till $10^6$ yrs before being constant while [SO$_2$/SO] and [SO/CS] ratios increase to a peak respectively at $5 times 10^4$ and $10^5$ yrs before rapidly decreasing till reaching a constant value. The ratios derived from our observations are not compatible with \citet{chapman2008} because they predict SO to be always more abundant than CS, except for objects younger than $10^3$ yrs, when the ratio can be less or equal to 1, but in that case the [SO$_2$/SO] ratio should be less than 10$^{-1}$ while the ratio is larger by an order of magnitude in our sample. 

From our work, the total abundance of sulphur (CS+SO+SO$_2$+H$_2$S+OCS)
can be estimated to be $2~10^{-9}-5~10^{-8}$, several orders of magnitude
lower than the corresponding solar abundance (3.4~$10^{-5}$) and the
value derived in the low-mass hot core IRAS16293 by
\citet{wakelam2004a} (2.8~$10^{-6}$). It is likely that we do not probe the very inner parts of the protostars where the S-bearing molecules are fully evaporated from the grain mantles. Hypothesis to explain such depletion of sulphur (including the one previously mentioned) will be explored in another paper on the chemistry.

%{\ bf Actually, we think that C$^{34}$S is not less abundant in W43MM1 but the isotopic lines we have observed do not probe the same layers %than the main isotopologue lines: there is probably a real abundance jump that is revealed by the C$^{34}$S lines.}

\section{Conclusion}

We have presented here new observations of sulfur-bearing species
toward $4$ massive dense cores. Different gas components, revealing
outflows or multiplicity, are detected in IRAS05358+3543,
IRAS18264$-$1152 and IRAS18162$-$2048. A strong infall of more than
2~\kms \ towards W43MM1 might be observed, leading to an impressive
kinematic mass infall rate. W43MM1 study shows that hot core may
appear earlier than expected in the evolution.

An evolutionary classification derived from the calculated SEDs is
proposed,  from W43MM1 to IRAS18162$-$2048, and is likely to be
observed in the molecular abundances too. More precisely, molecular
ratios like [OCS/H$_2$S], [CS/H$_2$S],  [SO/OCS], [SO$_2$/OCS],
[CS/SO] and [SO$_2$/SO], using low energy transitions, might be good
indicators of evolution depending on layers probed by the observed
molecular transitions: obvious trends along the massive core evolution
are seen for colder outer regions at 60~K, but drawing definitive
conclusions is more difficult for the inner regions. Observations of
molecular emission from warmer layers, hence implying higher upper
energy levels are mandatory to include. A follow-up with APEX
telescope should be made to have access to these higher frequencies.
Of course, the source multiplicity due the distance is problematic and
only ALMA will provide a definitive answer.

Nevertheless, a specific chemical modelling will be done for these sources in a forthcoming paper (Wakelam et al., in preparation) in order to explain the observed molecular ratios and to try to date the studied objects.

%\begin{acknowledgements}
    
%\end{acknowledgements}

%________________________________________________________________

\bibliography{biblio}
\bibliographystyle{aa}
\Online

\onlfig{14}{
   \begin{figure*}
   \centering
   \includegraphics[width=\columnwidth]{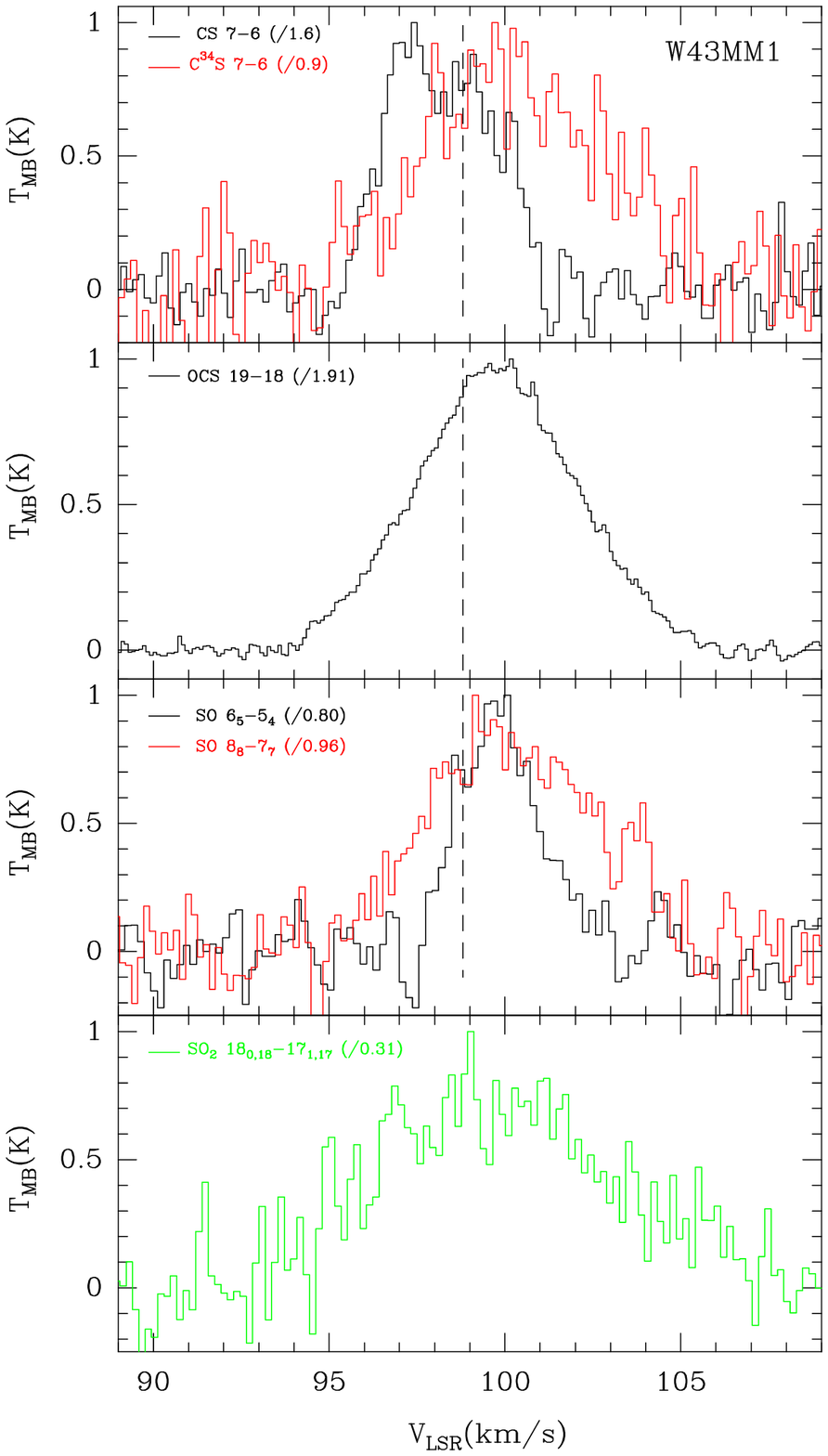}
   \caption{Normalized emission of CS, H$_2$S, OCS, SO, SO$_2$ (and isotopic species) lines from W43MM1. Spectra velocity resolutions are 0.10-0.19~\kms. The dashed line shows the source LSR velocity.}
    \end{figure*}
}

%\newpage

\onlfig{15}{
   \begin{figure*}
   \centering
   \includegraphics[width=\columnwidth]{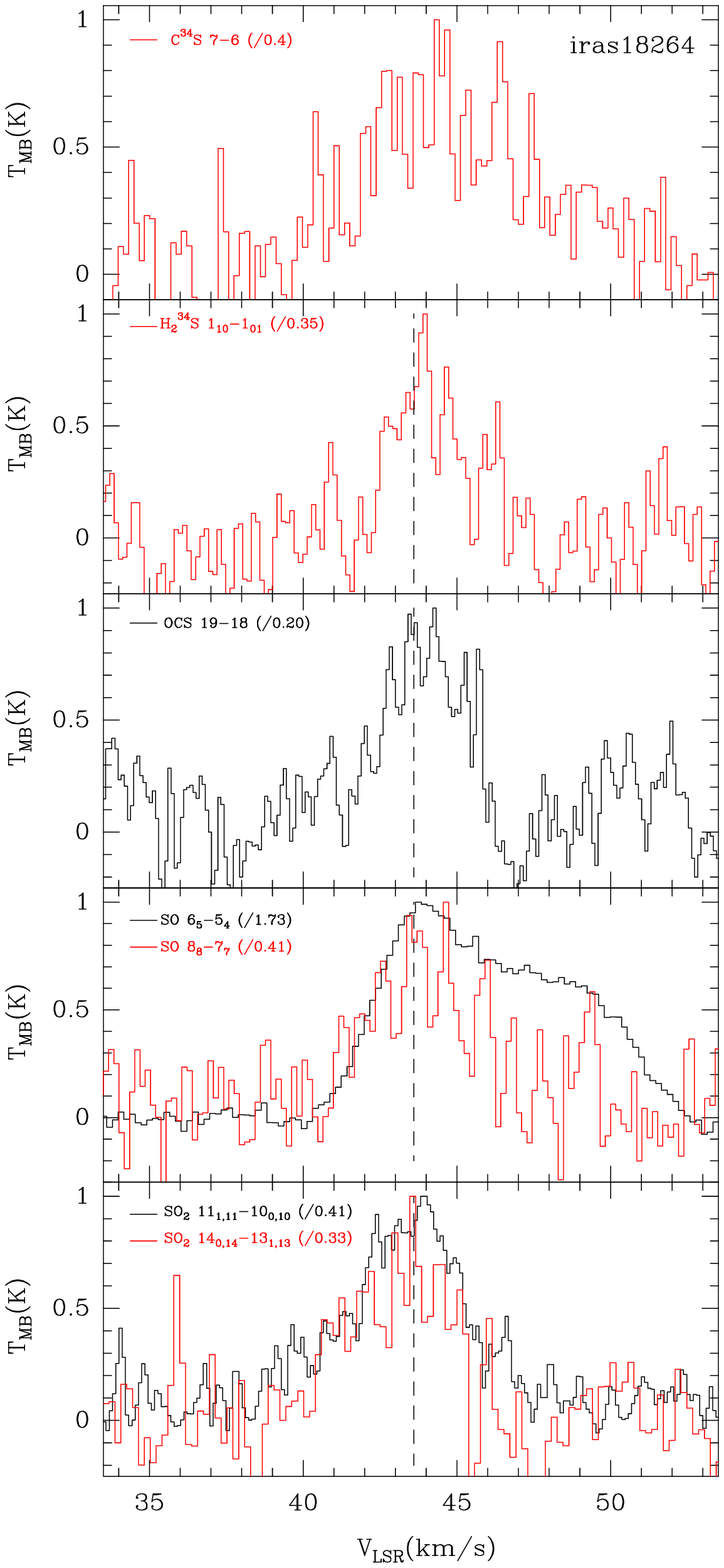}
   \caption{Normalized emission of CS, H$_2$S, OCS, SO, SO$_2$ (and isotopic species) lines from IRAS18264$-$1152. Spectra velocity resolutions are 0.10-0.19~\kms. The dashed line shows the source LSR velocity.}
    \end{figure*}
}

%\newpage

\onlfig{16}{
   \begin{figure*}
   \centering
   \includegraphics[width=\columnwidth]{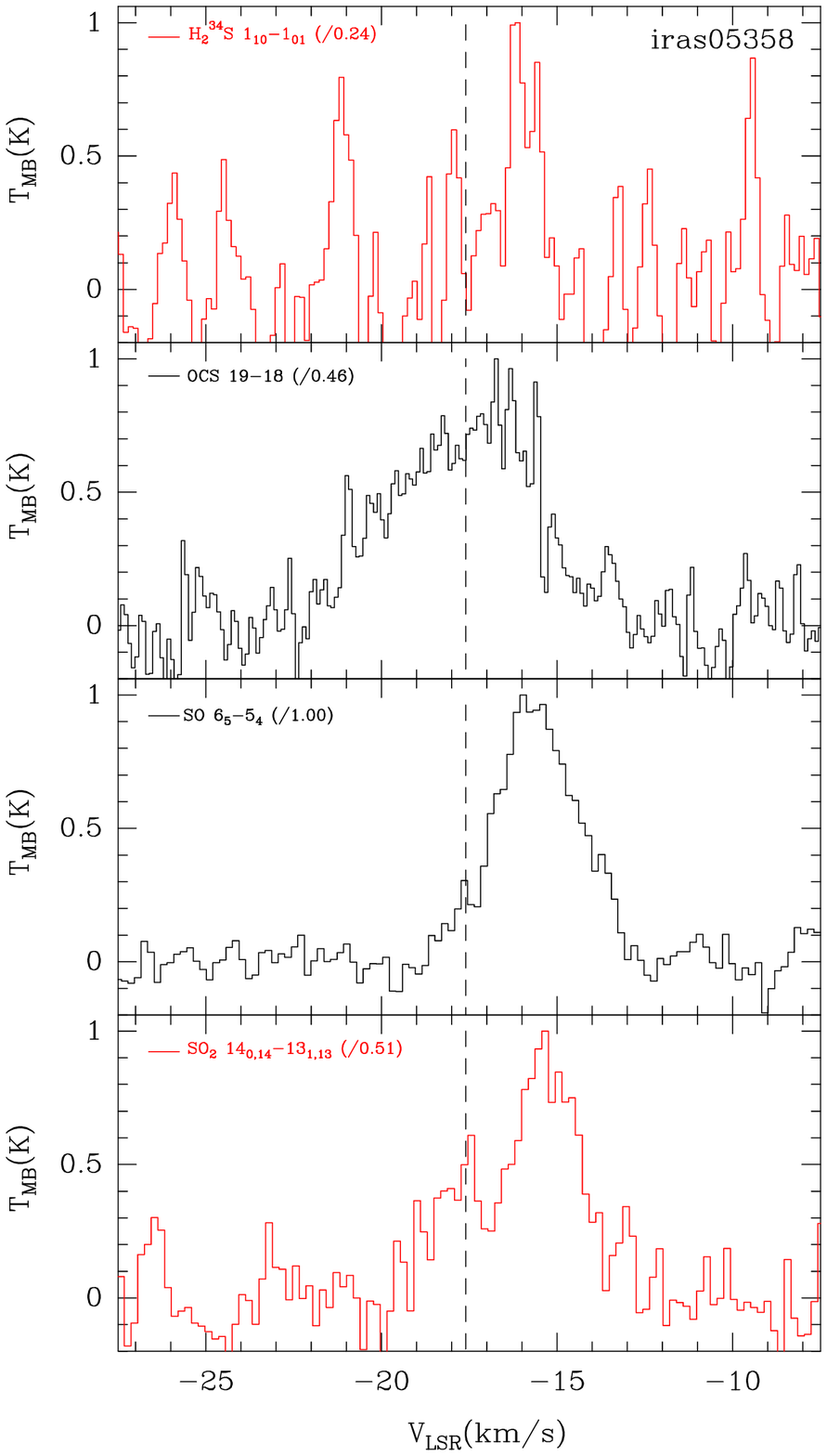}
   \caption{Normalized emission of CS, H$_2$S, OCS, SO, SO$_2$ (and isotopic species) lines from IRAS05358+3543. Spectra velocity resolutions are 0.10-0.19~\kms. The dashed line shows the source LSR velocity.}
    \end{figure*}
}

\onlfig{17}{
   \begin{figure*}
   \centering
   \includegraphics[width=\columnwidth]{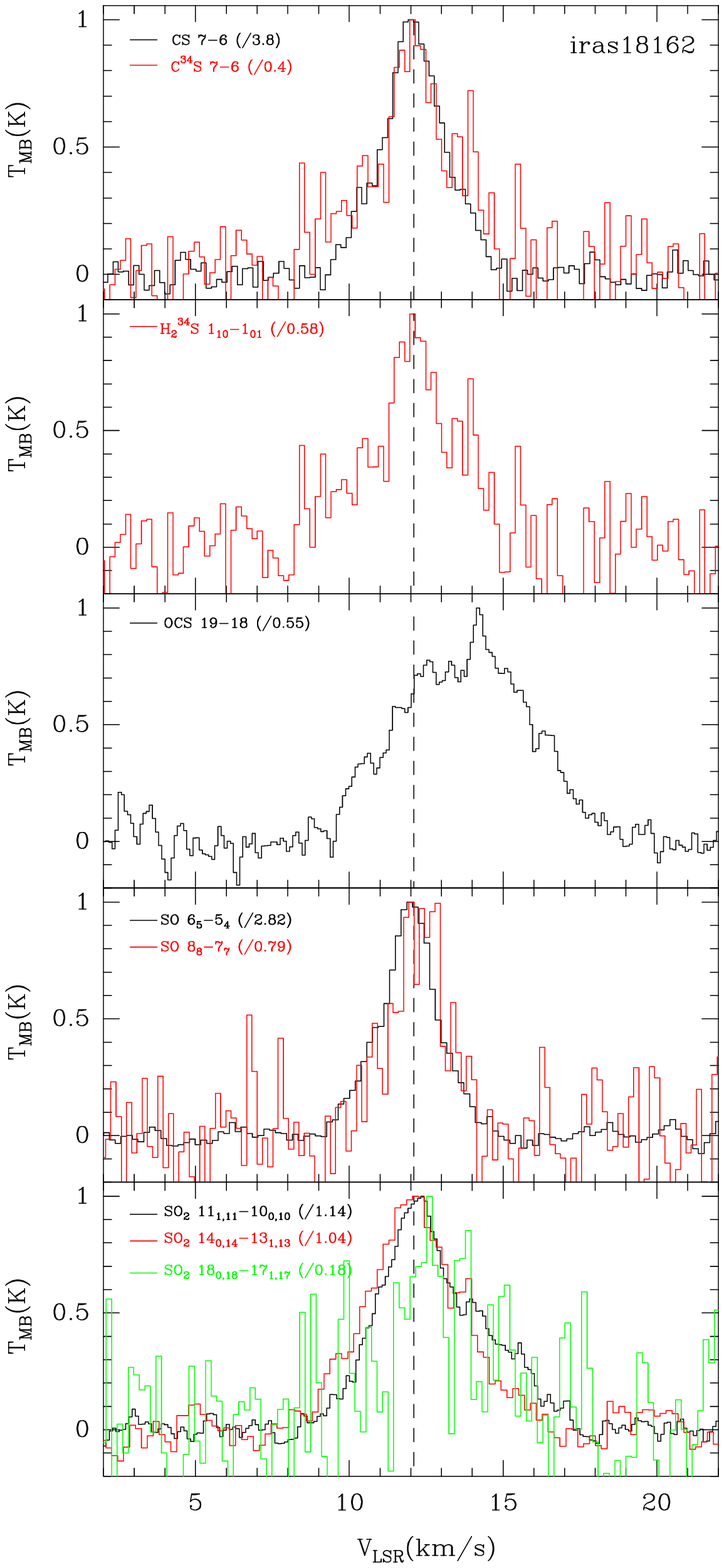}
   \caption{Normalized emission of CS, H$_2$S, OCS, SO, SO$_2$ (and isotopic species) lines from IRAS18162$-$2048. Spectra velocity resolutions are 0.10-0.19~\kms. The dashed line shows the source LSR velocity.}
    \end{figure*}
}

\onltab{6}{
\begin{table*}[t!]
\begin{minipage}[t]{500pt}
  \caption{Observed line emissions parameters for W43MM1. Widths ($\Delta\varv_{3\sigma}$ is the linewidths at 3$\sigma$ level;  $\Delta\varv_{1/2}$ is the half power linewidth) and line fluxes are derived from gaussian fits (several components) made with CLASS software. For the non detection, 1$\sigma$ upper limits are given.}\label{line-W43MM1}
\begin{center}
\begin{tabular}{lccccccccc} \hline \hline
%  \multicolumn{10}{c}{{\bf IRAS05358}}  \\ \hline
{\bf Species} & Transition  & $\Delta\varv$ & T$_\mathrm{mb}$ & $\sigma$ & $\int$T$_\mathrm{mb}$$\delta\varv$ & $\Delta\varv_{3\sigma}$ & $\Delta\varv_{1/2}$ & comment \\ 
 & & km$/$s & K & mK & K.km$/$s &  km$/$s &  km$/$s & \\ \hline
CS & 3-2  & 9.3(0.2) & 4.01 & 36 & 22.90(0.15) & 15.69 & 9.21 &  $^a$ \\
 & 5-4  & 9.68(0.04) & 6.85 & 62 & 61.6(0.3) & 14.50 & 9.36 &  $^a$ \\
 & 7-6  & 4.2(0.3) & 1.2 & 255 & 5.4(0.7) & 4.5 & 4.5 & $^b$ \\ \hline
C$^{34}$S & 3-2  & 5.69(0.02) & 2.12 & 42 & 12.96(0.06) & 11.23 & 6.02 &  \\
 & 7-6  & 6.1(0.3) & 1.05 & 269 & 4.8(0.2) & 2.7 & 5.6 &  \\ \hline
SO & $3_4-2_3$ & 6.10(0.01) & 2.66 & 23 & 17.52(0.03) & 13.24 & 6.62 &  \\
 & $5_6-4_5$  & 4.40(0.06) & 1.14 & 65 & 5.10(0.07) & 7.22 & 4.51 &  \\
 &$6_5-5_4$ & 2.6(0.1) & 0.8 & 85 & 2.01(0.09) & 3.47 & 2.54 &  \\
 & $8_8-7_7$  & 5.4(0.3) & 0.6 & 184 & 2.8(0.1) & 5.4 & 5.4 & weak det. \\ \hline
$^{34}$SO & $3_4-2_3$  & 5.6(0.1) & 0.39 & 26 & 2.42(0.04) & 10.22 & 5.52 &  \\  
 & $8_8-7_7$   &  &  & 197 & $<0.009$ &  & & no det. \\ \hline
SO$_2$ 	& $5_{1,5}-4_{0,4}$  & 5.05(0.1) & 0.52 & 38 & 2.33(0.05) & 8.63 & 3.91 &  \\
 			& $10_{0,10}-9_{1,9}$  & 5.1(0.3) & 0.33 & 43 & 1.41(0.06) & 8.02 & 4.0 &  \\
 			& $11_{1,11}-10_{0,10}$   &  &  & 47 & $<$0.006 &  &  & no det. \\
 	& $14_{0,14}-13_{1,13}$   &  &  & 60 & $<$0.01 &  &  & no det. \\
 			& $18_{0,18}-17_{1,17}$ & 7.8(0.3) & 0.23 & 80 & 1.89(0.07) & 7.8 & 7.8 &  \\
			& $28_{4,24}-28_{3,25}$   &  &  & 60 & $<$0.01 &  &  & no det. \\ \hline
$^{34}$SO$_2$ 	& $5_{1,5}-4_{0,4}$   & & & 32 & $<$0.006 & & & no det. \\  \hline
OCS 			& $8-7$   & 6.51(0.02) & 0.90 & 11 & 6.20(0.02) & 13.10 & 6.55 & $^a$ \\
 			& $13-12$  & 4.57(0.03) & 1.58 & 39 & 7.33(0.05) & 8.72 & 4.51 &  \\
 			& $19-18$  & 5.37(0.02) & 1.92 & 35 & 10.46(0.04) & 11.03 & 5.01 &  \\  \hline
OC$^{34}$S 		& $8-7$  & 7.26(0.03) & 0.14 & 7 & 1.13(0.01) & 13.63 & 6.92 &   \\ \hline
H$_2$S 	& $1_{1,0}-1_{0,1}$  & 9.1(0.2) & 1.06 & 72 & 12.52(0.6)$^e$  & 12.50 & 2.76 & $^a$  \\
 	& $2_{2,0}-2_{1,1}$   & 4.15(0.04) & 1.26 & 45 & 5.40(0.05) & 7.62 & 4.11 &   \\ \hline
H$_2$$^{34}$S & $1_{1,0}-1_{0,1}$   &  &  & 72 & $<$0.01 &  &  &  no det. \\
\hline 
\end{tabular}
\end{center}
$^a$ self-absorption at $\varv_\mathrm{source}$ \\
$^b$ line blend with H$_2$CO line from image band. \\
\end{minipage}
\end{table*}
}

\onltab{7}{
\begin{table*}[t!]
\begin{minipage}[t]{500pt}
  \caption{Observed line emissions parameters for IRAS18264$-$1152. Widths ($\Delta\varv_{3\sigma}$ is the linewidths at 3$\sigma$ level;  $\Delta\varv_{1/2}$ is the half power linewidth) and line fluxes are derived from gaussian fits (several components) made with CLASS software. For the non detection, 1$\sigma$ upper limits are given.}\label{line-iras18264 }
\begin{center}
\begin{tabular}{lccccccccc} \hline \hline
%  \multicolumn{10}{c}{{\bf IRAS05358}}  \\ \hline
{\bf Species} & Transition  & $\Delta\varv$ & T$_\mathrm{mb}$ & $\sigma$ & $\int$T$_\mathrm{mb}$$\delta$v & $\Delta\varv_{3\sigma}$ & $\Delta\varv_{1/2}$ & comment \\ 
 &  & km$/$s & K & mK & K.km$/$s &  km$/$s &  km$/$s & \\ \hline
CS & 3-2 ) & 4.19(0.01) & 7.2 & 54 & 31.49(0.07) & 10.28 & 4.01 &   \\
 & 5-4  & 2.97(0.03) & 2.26 & 45 & 6.81(0.05) & 7.32 & 2.81 &   \\
 & 7-6  & 3.2(0.8)? & 0.6? & 500 & 1.9(0.5)? & ? & ? & $^b$   \\ \hline
C$^{34}$S & 3-2  & 3.67(0.04) & 1.64 & 42 & 6.02(0.05) & 8.32 & 3.41 &  \\
 & 7-6   & 5.1(0.5) & 0.34 & 111 & 1.1(0.1) & 3.7 & 4.7 & tentative det. \\ \hline
SO & $3_4-2_3$  & 3.27(0.02) & 3.11 & 44 & 10.38(0.05) & 7.32 & 3.21 &  \\
 & $5_6-4_5$  & 2.98(0.02) & 2.33 & 53 & 7.08(0.05) & 6.42 & 2.81 &  \\
 &$6_5-5_4$  & 6.2(0.1) & 1.72 & 45 & 11.8(0.2) & 11.36 & 7.62 & $^a$ \\
 & $8_8-7_7$  & 4.1(0.4) & 0.44 & 116 & 1.3(0.1) & 2.2 & 4.8 & weak det. \\ \hline
$^{34}$SO & $3_4-2_3$  & 3.02(0.2) & 0.19 & 39 & 0.60(0.04) & 4.01 & 4.01 &  $^d$ \\   
 &$8_8-7_7$ &    &  & 132 &  $<0.006$ &  & & no det. \\ \hline
SO$_2$ 	& $5_{1,5}-4_{0,4}$  & 3.30(0.3) & 0.24 & 33 & 0.91(0.05) & 5.91 & 5.59 & \\
 			& $10_{0,10}-9_{1,9}$  & 2.9(0.2) & 0.19 & 34 & 0.50(0.03) & 3.03 & 3.03 &  weak line \\
 			& $11_{1,11}-10_{0,10}$  & 4.1(0.1) & 0.40 & 41 & 1.56(0.04) & 7.02 & 3.31 & \\
 	& $14_{0,14}-13_{1,13}$  & 3.9(0.3) & 0.34 & 59 & 0.99(0.07) & 3.95 & 3.95 & \\
 			& $18_{0,18}-17_{1,17}$   &  &  & 79 & $<0.004$ &  & & no det. \\ 
			& $28_{4,24}-28_{3,25}$   & &  & 63 & $<$0.01 &  &  & no det. \\ \hline
$^{34}$SO$_2$ 	& $5_{1,5}-4_{0,4}$   & & & 64 & $<$0.01 & & & no det. \\  \hline
OCS 			& $8-7$   & 3.73(0.1) & 0.18 & 5 & 0.71(0.01) & 9.19 & 3.51 &  \\
 			& $13-12$ & 3.3(0.1) & 0.41 & 46 & 1.35(0.05) & 4.13 & 3.58 &  \\
 			& $19-18$  & 2.7(0.1) & 0.20 & 34 & 0.49(0.03) & 3.21 & 3.21 & bad data \\    \hline
OC$^{34}$S 		& $8-7$   &  &  & 4 & $<$0.001 & & & no det. \\ \hline
H$_2$S 	& $1_{1,0}-1_{0,1}$  & 3.63(0.04) & 1.94 & 59 & 7.32(0.06) & 7.22 & 3.61 &   \\
 	& $2_{2,0}-2_{1,1}$   & 2.5(0.3) & 0.14 & 28 & 0.33(0.03) & 5.51 & 2.81 & $^c$  \\ \hline
H$_2$$^{34}$S & $1_{1,0}-1_{0,1}$  & 2.7(0.1) & 0.35 & 60 & 0.68(0.06) & 3.72 & 3.72 &   \\
\hline 
\end{tabular}
\end{center}
$^a$ self-absorption at  $\varv_\mathrm{source}$ \\
$^b$ line blend with H$_2$CO line from image band. \\
$^c$ unidentified line @ 34.87~km$/$s (216.716~GHz).\\
$^d$ + other line @ 29.1~\kms\ (135.78223~GHz), T$_\mathrm{mb}=0.14$~K,  $\delta\varv=1.1$~\kms, $\int$T$_\mathrm{mb}$$\delta\varv=0.17$~\kkms \\
\end{minipage}
\end{table*}
}

\onltab{8}{
\begin{table*}[t!]
\begin{minipage}[t]{500pt}
  \caption{Observed line emissions parameters for IRAS05358$+$3543. Widths ($\Delta\varv_{3\sigma}$ is the linewidths at 3$\sigma$ level;  $\Delta\varv_{1/2}$ is the half power linewidth) and line fluxes are derived from gaussian fits (several components) made with CLASS software. For the non detection, 1$\sigma$ upper limits are given.}\label{line-iras05358 }
\begin{center}
\begin{tabular}{lcccccccc} \hline \hline
%  \multicolumn{10}{c}{{\bf IRAS05358}}  \\ \hline
{\bf Species} & Transition  & $\Delta\varv$ & T$_\mathrm{mb}$ & $\sigma$ & $\int$T$_\mathrm{mb}$$\delta\varv$ & $\Delta\varv_{3\sigma}$ & $\Delta\varv_{1/2}$ & comment \\ 
 &  & km$/$s & K & mK & K.km$/$s &  km$/$s &  km$/$s & \\ \hline
CS & 3-2   & 5.54(0.06) & 7.92 & 35 & 29.75(0.02) & 11.07 & 2.77 & $^a$ \\
 &5-4  & 3.60(0.10) & 5.14 & 59 & 16.64(0.07) & 7.37 & 3.31 & $^a$ \\
 & 7-6   &  &  & &  &  &  & no data \\ \hline
C$^{34}$S & 3-2  & 4.30(0.16) & 1.59 & 24 & 4.10(0.01) & 7.12 & 2.51 & $^a$  \\
 & 7-6 &    &  & &  &  &  & no data \\  \hline
SO & $3_4-2_3$  & 3.97(0.05) & 2.42 & 45 & 11.46(0.01) & 8.42 & 4.71 & $^a$  \\
 & $5_6-4_5$  & 3.31(0.05) & 2.29 & 63 & 7.09(0.06) & 6.62 & 2.91 & $^a$ \\
 &$6_5-5_4$  & 2.81(0.08) & 0.99 & 60 & 2.82(0.06) & 4.54 & 2.81 &  $^a$\\
 & $8_8-7_7$   &  &  & &  &  &  & no data \\  \hline
$^{34}$SO & $3_4-2_3$  & 4.47(0.35) & 0.26 & 50 & 1.1 &  4.13 & 4.4 &$^a$ \\   
 &$8_8-7_7$   &  &  & &  &  &  & no data \\  \hline
SO$_2$ 	& $5_{1,5}-4_{0,4}$  & 4.9(0.2) & 0.31 & 36 & 1.48(0.05) & 5.91 & 4.55  & \\
 			& $10_{0,10}-9_{1,9}$  & 3.9(0.4) & 0.24 & 49 & 0.70(0.06) & 1.90 & 4.81 &  \\
 			& $11_{1,11}-10_{0,10}$ &    &  & 55 & $<$0.01 &  &  & no det. \\
 	& $14_{0,14}-13_{1,13}$ & 3.89(0.3) & 0.50 & 68 & 1.49(0.09) & 4.11 & 3.41 &  \\
 			& $18_{0,18}-17_{1,17}$&    &  & &  &  &  & no data \\ 
			& $28_{4,24}-28_{3,25}$ &   & & 45 & $<$0.008 &  &  & no det. \\  \hline
$^{34}$SO$_2$ 	& $5_{1,5}-4_{0,4}$   & & & 25 & $<$0.005 & & & no det. \\  \hline
OCS 			& $8-7$   & 4.69(0.05) & 0.27 & 10 & 1.29(0.01) & 12.9 & 5.01 & $^a$ \\
 			& $13-12$  & 3.08(0.05) & 0.20 & 27 & 0.58(0.05) & 3.11 & 3.01 & $^a$ \\
 			& $19-18$ & 5.2(0.2) & 0.46 & 58 & 2.01(0.06) & 5.61 & 5.51 & $^a$ \\  \hline
OC$^{34}$S 		& $8-7$&    &  & 10 & $<$0.003 & &  & no det. \\  \hline
H$_2$S 	& $1_{1,0}-1_{0,1}$ & 4.8(0.15) & 2.56 & 53 & 9.36(0.15) & 8.46 & 3.03 & $^a$ \\
 	& $2_{2,0}-2_{1,1}$   & 2.98(0.15) & 0.37 & 44 & 0.89(0.04) & 3.91 & 3.41 &  $^a$ \\  \hline
H$_2$$^{34}$S & $1_{1,0}-1_{0,1}$  & 1.2(0.3) & 0.24 & 77 & 0.25(0.05) & 0.83 & 1.01 & very weak line$^b$ \\
\hline 
\end{tabular}
\end{center}
$^a$ self-absorption at  $\varv_\mathrm{source}$ \\
$^b$ detected at 2$\sigma$ with two peaks @ $-15.57$ and $-16.21$~km$/$s \\
\end{minipage}
\end{table*}
}

\onltab{9}{
\begin{table*}[t!]
\begin{minipage}[t]{500pt}
  \caption{Observed line emissions parameters for IRAS18162$-$2048. Widths($\Delta\varv_{3\sigma}$ is the linewidths at 3$\sigma$ level;  $\Delta\varv_{1/2}$ is the half power linewidth) and line fluxes are derived from gaussian fits (several components) made with CLASS software. For the non detection, 1$\sigma$ upper limits are given.}\label{line-iras18162}
\begin{center}
\begin{tabular}{lcccccccc} \hline \hline
%  \multicolumn{10}{c}{{\bf IRAS05358}}  \\ \hline
{\bf Species} & Transition  & $\Delta\varv$ & T$_\mathrm{mb}$ & $\sigma$ & $\int$T$_\mathrm{mb}$$\delta\varv$ & $\Delta\varv_{3\sigma}$ & $\Delta\varv_{1/2}$ & comment \\ 
 & & km$/$s & K & mK & K.km$/$s &  km$/$s &  km$/$s & \\ \hline
CS & 3-2   & & &  &  & & & no obs. \\
 &5-4  & 2.93(0.01) & 7.41 & 38 & 22.10(0.03) & 6.78 & 2.91 &   \\
 & 7-6  & 2.46(0.04) & 4.05 & 237 & 9.3(0.1) & 4.08 & 2.14 &  \\ \hline
C$^{34}$S & 3-2   & & &  &  & & & no obs. \\
 & 7-6   & 2.1(0.3) & 0.47 & 116 & 0.87(0.06) & 1.17 & 2.63 &  \\ \hline
SO & $3_4-2_3$  & 2.24(0.01) & 5.27 & 40 & 12.00(0.04) & 6.46 & 2.10 &   \\
 & $5_6-4_5$  & 2.20(0.01) & 5.22 & 45 & 11.31(0.03) & 5.42 & 4.84 &   \\
 &$6_5-5_4$  & 2.24(0.03) & 2.80 & 79 & 6.25(0.08) & 4.52 & 1.78 &   \\
 & $8_8-7_7$  & 2.2(0.2) & 1.04 & 230 & 1.7(0.1) & 1.1 & 1.7 &  \\ \hline
$^{34}$SO & $3_4-2_3$   & & & 38 & $<$0.007 & & & no det. \\   
 &$8_8-7_7$ &   &  & 123 &  $<0.006$ &  &  & no det.  \\  \hline
SO$_2$ 	& $5_{1,5}-4_{0,4}$  & 3.3(0.1) & 0.72 & 34 & 1.86(0.04) & 6.46 & 1.68 &  \\
 			& $10_{0,10}-9_{1,9}$  & 1.69(0.07) & 0.43 & 33 & 0.75(0.02) & 2.39 & 1.75 & $^a$  \\
 			& $11_{1,11}-10_{0,10}$ ) & 3.86(0.08) & 1.14 & 35 & 4.20(0.07) & 7.88 & 3.49 &   \\
 	& $14_{0,14}-13_{1,13}$  & 3.63(0.09) & 1.02 & 59 & 3.71(0.07) & 5.81 & 3.68 &  \\
 			& $18_{0,18}-17_{1,17}$  & 1.2(0.5) & 0.13 & 75 & 0.16(0.05) &  & 3.7? & tentative det. \\ 
			& $28_{4,24}-28_{3,25}$   &  &  & 42 & $<$0.007 &  &  &  no det. \\ \hline
$^{34}$SO$_2$ 	& $5_{1,5}-4_{0,4}$    & & & 38 & $<$0.007 & & & no det. \\  \hline
OCS 			& $8-7$   & 3.5(0.1) & 0.12 & 8 & 0.40(0.01) & 7.17 & 3.10 &   \\
 			& $13-12$  & 4.7(0.5) & 0.23 & 51 & 0.75(0.06) & 2.85 & 6.26 &   \\
 			& $19-18$  & 5.12(0.08) & 0.55 & 33 & 2.53(0.04) & 7.55 & 4.55 &   \\  \hline
OC$^{34}$S 		& $8-7$&   & & 9 & $<$0.003 &  & & no det.  \\ \hline
H$_2$S 	& $1_{1,0}-1_{0,1}$  & 3.46(0.02) & 2.92 & 50 & 10.00(0.05) & 8.0 & 3.36 & $^a$  \\
 	& $2_{2,0}-2_{1,1}$   & 3.2(0.3) & 0.24 & 24 & 0.78(0.07) & 4.26 & 2.97 & $^b$  \\  \hline
H$_2$$^{34}$S & $1_{1,0}-1_{0,1}$  & 2.3(0.1) & 0.60 & 65 & 1.15(0.06) & 2.58 & 1.55 &   \\
\hline 
\end{tabular}
\end{center}
$^a$ self-absorption at  $\varv_\mathrm{source}$ \\
$^b$ unidentified lines @ $216.701026$ and $216.71.6$~GHz ($25$ and $4$~km$/$s) \\
\end{minipage}
\end{table*}
}

\onltab{10}{
\begin{table*}[t!]
\begin{minipage}[t]{500pt}
  \caption{Opacities derived from the modelling of the molecular emission of the sources. "ND" stands for no data. }\label{opacities}
\begin{center}
\begin{tabular}{lcccc} \hline \hline
{\bf Lines} &  {\bf W43-MM1}	& {\bf 18264}	&	{\bf 05358}	&	{\bf 18162} \\ \hline
CS 3-2	&	52.0	&	1.6	&	0.45	&	ND	\\
CS 5-4	&	65.0	&	0.31	&	0.81	&	0.67 \\
CS 7-6	&	300	&	0.94	&	ND	&	2.8 \\ \hline
C$^{34}$S 3-2	&	2.8	&	0.18	&	0.09	&	ND \\
C$^{34}$S 7-6	&	16.0	&	0.51	&	ND	&	0.41 \\ \hline
OCS 8-7	&	6.0	&	0.07	&	0.04	&	0.06 \\
OCS 13-12	&	2.6	&	0.1	&	0.06	&	0.1 \\	 	
OCS 19-18	&	4.9	&	0.12	&	0.07	&	0.2 \\ \hline
OC$^{34}$S 8-7	&	0.4	&	0.03	&	0.02 & 0.05 \\ \hline
H$_2$S  $1_{1,0}-1_{0,1}$ 	&	4.2	&	0.53	&	0.21	&	3.9 \\
H$_2$S $2_{2,0}-2_{1,1}$	&	3.3	&	0.13	&	0.08	&	0.15 \\ \hline
H$_2$$^{34}$S  $1_{1,0}-1_{0,1}$	&	0.36	&	0.11	&	0.04	&	0.48 \\ \hline
SO  $3_4-2_3$	&	6.3	&	0.25	&	0.32	&	0.36 \\
SO $5_6-4_5$ 	&	0.72	&	0.21	&	0.15	&	0.39 \\
SO $6_5-5_4$	&	0.52	&	0.21	&	0.10	&	0.28 \\
SO $8_8-7_7$	&	6.8	&	0.33	&	ND	&	0.59 \\ \hline
$^{34}$SO  $3_4-2_3$	&	0.45	&	0.05	&	0.11	&	0.89 \\
$^{34}$SO $8_8-7_7$ 	&	1.6	&	0.26	&	ND	&	2.9 \\ \hline
SO$_2$  $5_{1,5}-4_{0,4}$	&	0.42	&	0.06	&	0.03	&	0.08 \\
SO$_2$ $10_{0,10}-9_{1,9}$	&	0.33	&	0.06	&	0.03	&	0.04 \\
SO$_2$ $11_{1,11}-10_{0,10}$ 	&	ND	&	0.11	&	ND	&	0.13 \\
SO$_2$ $14_{0,14}-13_{1,13}$	&	ND	&	0.21	&	0.23	&	0.42 \\
SO$_2$ $18_{0,18}-17_{1,17}$	&	6.0	&	0.24	&	ND	&	0.22 \\ \hline
$^{34}$SO$_2$  $5_{1,5}-4_{0,4}$	&	0.18	&	0.61	&	0.02	&	0.03 \\
\hline 
\end{tabular}
\end{center}
\end{minipage}
\end{table*}
}

\
\end{document}